\documentclass[twocolumn,prc,floatfix,showpacs,preprintnumbers,nofootinbib,superscriptaddress]{revtex4}
\pdfoutput=1
\usepackage{mathrsfs}
\usepackage{amssymb}
\usepackage{amsmath}
\usepackage{graphicx}
\usepackage{slashed}
\usepackage{bbold}
\usepackage[normalem]{ulem}

\usepackage{xcolor}
\definecolor{lcolor}{rgb}{0.5,0,0}
\definecolor{citcolor}{rgb}{0,0.3,0.0}

\usepackage[breaklinks,colorlinks,urlcolor=blue,citecolor=citcolor,linkcolor=lcolor]{hyperref}

\newcommand{\rt}{{\mathbf{r}}}

\newcommand{\xt}{{\mathbf{x}}}
\newcommand{\bt}{{\mathbf{b}}}

\newcommand{\yt}{{\mathbf{y}}}

\newcommand{\ut}{{\mathbf{u}}}
\newcommand{\vt}{{\mathbf{v}}}

\newcommand{\pt}{{\mathbf{p}}}
\newcommand{\qt}{{\mathbf{q}}}
\newcommand{\kt}{{\mathbf{k}}}
\newcommand{\ot}{\mathbf{0}}

\newcommand{\ptt}{p_T} 
\newcommand{\qtt}{q_T} 

\newcommand{\nabt}{\boldsymbol{\nabla}_T}

\newcommand{\ud}{\, \mathrm{d}}

\newcommand{\tr}{\, \mathrm{Tr} \, }

\newcommand{\nc}{{N_\mathrm{c}}}

\newcommand{\cf}{C_\mathrm{F}}

\newcommand{\nr}[1]{(\ref{#1})}

\newcommand{\qs}{Q_\mathrm{s}}

\newcommand{\lqcd}{\Lambda_{\mathrm{QCD}}}

\newcommand{\fig}{Fig.~}

\newcommand{\eq}{Eq.~}

\newcommand{\eqs}{Eqs.~}

\newcommand{\id}{\mathbb{1}}

\newcommand{\A}{{\mathcal{A}}}

\newcommand{\ahat}{{\hat{a}}}

\begin{document}

\author{T. Lappi}
\affiliation{Department of Physics, P.O. Box 35, 40014 University of Jyv\"askyl\"a, Finland}
\affiliation{Helsinki Institute of Physics, P.O. Box 64, 00014 University of Helsinki, Finland}
\author{B. Schenke}
\affiliation{Physics Department, Brookhaven National Laboratory, Upton, NY 11973, USA}
\author{S. Schlichting}
\affiliation{Physics Department, Brookhaven National Laboratory, Upton, NY 11973, USA}
\author{R. Venugopalan}
\affiliation{Physics Department, Brookhaven National Laboratory, Upton, NY 11973, USA}
\affiliation{Institut f\"{u}r Theoretische Physik, Universit\"{a}t Heidelberg, Philosophenweg 16, 69120 Heidelberg, Germany}

\title{Tracing the origin of azimuthal gluon correlations in the color glass condensate}

\pacs{24.85.+p,25.75.-q,12.38.Mh, 12.38.Lg}

\preprint{}

\begin{abstract}
 We examine the origins of azimuthal correlations observed in high energy proton-nucleus collisions by considering the simple example of the scattering of uncorrelated partons off color fields in a large nucleus.  We demonstrate how the physics of fluctuating color fields in the color glass condensate (CGC) effective theory generates these azimuthal multiparticle correlations and compute the corresponding Fourier coefficients $v_{n}$ within different CGC approximation schemes.  We discuss in detail the qualitative and quantitative differences between the different schemes. We will show how a recently introduced color field domain model that captures key features of the observed azimuthal correlations can be understood in the CGC effective theory as a model of non-Gaussian correlations in the target nucleus.
\end{abstract}

\maketitle

\section{Introduction}

Azimuthal anisotropies of multiparticle correlations observed in small systems such as those produced in p+Pb, d+Au, or $^3$He+Au collisions have been computed in various theoretical frameworks. Within different calculations these correlations are either dominantly due to initial state parton correlations in the projectile and target~\cite{Dumitru:2010iy,Dusling:2012iga,Dusling:2012cg,Dusling:2012wy,Dusling:2013oia,Kovner:2010xk,Kovner:2011pe,Kovchegov:2012nd,Kovchegov:2013ewa,Dumitru:2014dra,Dumitru:2014vka,Gyulassy:2014cfa,Lappi:2015vha,Schenke:2015aqa,Altinoluk:2015uaa} or from final state correlations that are generated by the collective flow of matter produced in the collision~\cite{Avsar:2010rf,Bozek:2013uha,Bozek:2013ska,Bzdak:2013zma,Qin:2013bha,Schenke:2014zha,Bzdak:2014dia}. Since both of these frameworks are able to describe key features of the data, disentangling the different effects and understanding how correlations are generated in both approaches is essential to obtain novel insight into the QCD dynamics of ultradense parton systems.

We will focus in this work on multiparticle correlations which are generated in the initial state. We will discuss within the color glass condensate (CGC) effective theory of high energy QCD~\cite{Gelis:2010nm} the origin of these correlations and we will critically examine the assumptions underlying different calculations in this framework. Even though some of the models may appear very different, their common features and their differences can be understood systematically as approximations to the underlying QCD dynamics.

We will orient our discussion of initial state correlations within the ``dilute-dense'' power counting in the CGC, where the incoming parton densities are assumed to be small in the projectile but large in the target nucleus. In this limit, analytical and numerical computations are comparatively simple and thus permit systematic comparisons between different approximation schemes. We note however that the kinematic region where the strongest azimuthal correlations are seen in experiments correspond more to a ``dense-dense'' situation as the parton densities are also large in the incoming projectile. A systematic power counting then requires one to solve classical Yang-Mills equations in the presence of projectile and target color sources, which
has only been achieved numerically~\cite{Lappi:2009xa,Schenke:2015aqa}. While there are important qualitative differences between dense-dense and dilute-dense systems, we believe that the lessons inferred from analytical and numerical studies the dilute-dense case can nevertheless be valuable for the discussion of the phenomenologically more relevant dense-dense collision systems.
 
We will here compute the two particle correlation function for quarks scattering off a large nucleus in the dilute-dense CGC description. The main ingredient of this computation is the so-called ``dipole-dipole'' correlator of light-like Wilson lines and we will compute this quantity in the following approximations: i) the Gaussian two gluon exchange  (``Glasma graph'') approximation \cite{Dumitru:2008wn,Gavin:2008ev,Dusling:2012cg,Dusling:2009ar,Gelis:2009wh,Dusling:2009ni,Dumitru:2010iy,Dusling:2012iga,Dusling:2012wy,Dusling:2013oia} and ii)  the nonlinear Gaussian approximation \cite{Iancu:2002aq,Blaizot:2004wu,Blaizot:2004wv,Weigert:2005us,Kovchegov:2008mk,Fujii:2006ab,Marquet:2010cf,Dumitru:2011vk,Dominguez:2011wm,Iancu:2011ns,Iancu:2011nj}. We will determine the second, third, and fourth azimuthal Fourier coefficients of the two particle correlation function within these two approximation schemes and compare the results to numerical lattice simulations of the full correlator in the McLerran--Venugopalan (MV) model~\cite{McLerran:1994ni,McLerran:1994ka,McLerran:1994vd} and after renormalization group (JIMWLK) evolution~\cite{JalilianMarian:1997gr,JalilianMarian:1997dw,Iancu:2000hn,Ferreiro:2001qy} of the MV model initial conditions to higher rapidities~\cite{Lappi:2015vha}. We will further discuss how our computations relate  to the ``color field domain model'' introduced in  \cite{Dumitru:2014dra,Dumitru:2014yza,Dumitru:2014vka,Skokov:2014tka} based on ideas developed in \cite{Kovner:2010xk,Kovner:2011pe}. Since many features of this model appear similar to those discussed previously~\cite{Dumitru:2010iy,Dusling:2012iga,Dusling:2012wy,Dusling:2013oia}, it is important to understand the interpretation and justification for this model from first principles. We will demonstrate that the effects of the color field domain model can be reproduced in the CGC effective theory if non-Gaussian correlations are assumed to play an important role.

This paper is organized as follows. In the next section, we shall discuss the physical picture of how initial state multiparticle correlations are generated and derive the leading order expressions for single and double inclusive distributions. These can be expressed in terms of correlators of lightlike Wilson lines, which we will  calculate in Sec. 3 within the Glasma graph and nonlinear Gaussian approximations. We then compute the Fourier moments $v_n$ of two particle correlations in both approximation schemes in Sec. 4 and compare our results with numerical lattice simulations of the full correlation function. In Section 5 the analytical and numerical results obtained are then compared to the color field domain model. We first cast our analytical results in terms of color electric field correlators and make a direct comparison with expressions for the same correlator in the color field domain model.  We will demonstrate that our results  provide a clear interpretation of the color field domain model which clarifies the discussion in the recent literature.  We end with a summary of the results of the paper and an outlook on further research directions in computations of multiparton correlations in high energy QCD.
 
\begin{figure}[t!]
   \begin{center}   
     	
       \includegraphics[width=0.3\textwidth]{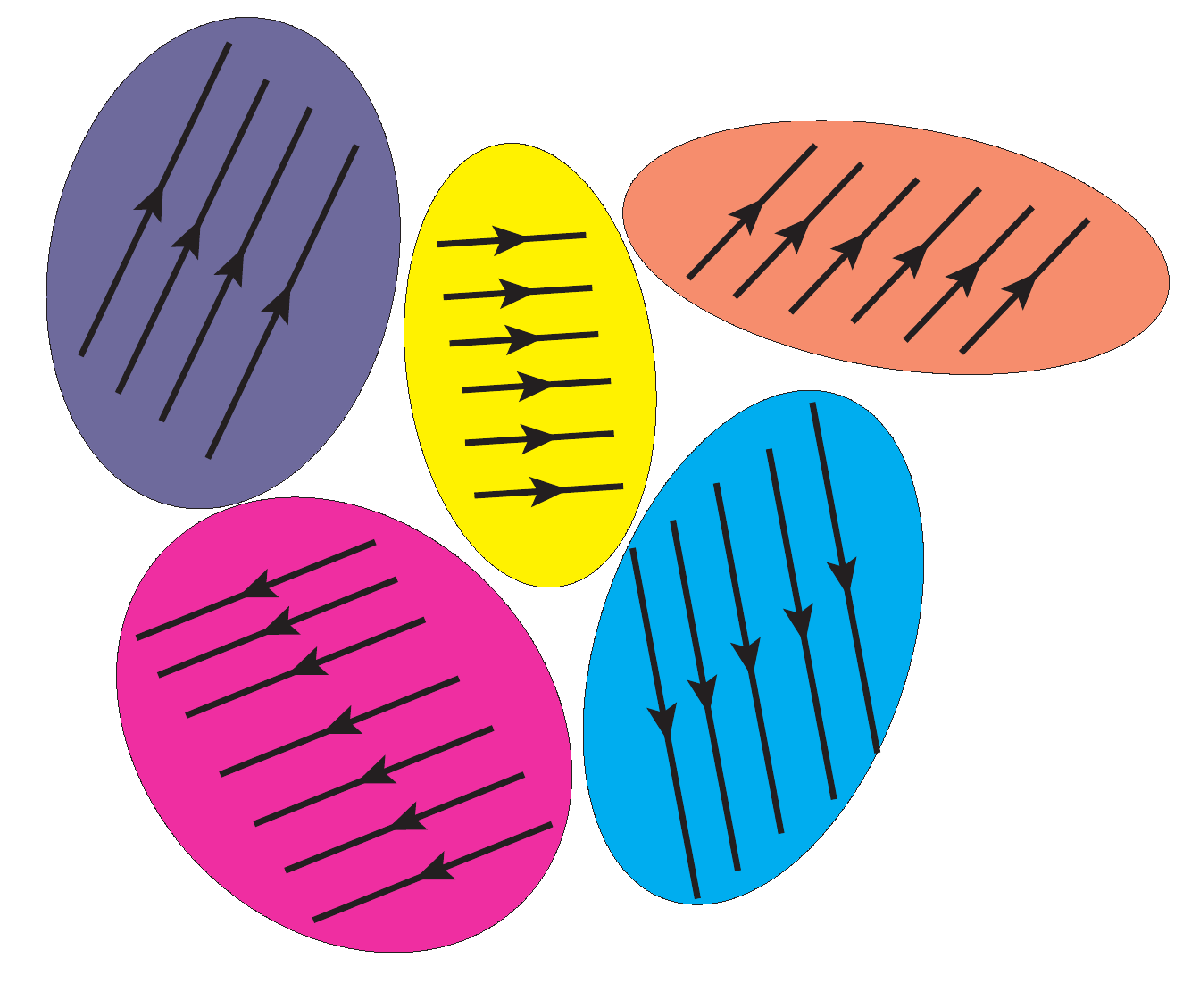} 

     \caption{ \label{fig:ColorFields} (Color online) Color electric fields inside the nucleus fluctuate on an event by event basis.}
      \end{center}
\end{figure}

\section{Multiparticle correlations from fluctuating color fields}
\label{sec:hybridcorr}

We begin our discussion of the physics of initial state correlations with the simplest possible example of the high energy scattering of individual (uncorrelated) quarks off a large nucleus. Our general picture is that each parton scatters independently off the color field of the nucleus receiving a transverse momentum kick in the process.  As noted previously~\cite{Dumitru:2008wn,Kovner:2010xk,Kovner:2011pe}, the color fields fluctuate from event to event and are locally organized in domains of size $\sim 1/\qs$ as illustrated in \fig\ref{fig:ColorFields}. When two (or more) quarks scatter off the same domain, they will receive a similar kick whenever they are in the same color state. This leads to a correlation which is suppressed by $1/\nc^2$ (in the limit of large $\nc$) and the number of domains $\qs^2S_\bot$, where $S_{\bot}$ denotes the transverse area  probed by the projectile. We will now discuss this physical picture in in more detail and further develop its quantitative implementation along the lines of the discussion in Ref.~\cite{Lappi:2015vha}.

\subsection{Single quark scattering}
Within the CGC formalism, the color fields inside the target nucleus are determined by the solution of the classical Yang-Mills equations 
\begin{eqnarray}
[D_{\mu},F^{\mu\nu}]=J^{\nu}\;,
\end{eqnarray}
where the eikonal current $J^{\mu}$ is given in terms of the density of color charges $\rho$ inside the target nucleus as
\begin{eqnarray}
J^{\mu}(\xt,x^{+})=\delta^{\mu-}\rho(\xt,x^{+})\;.
\end{eqnarray}
The solution to the classical Yang-Mills equations takes the well known 
form~\cite{Iancu:2003xm}
\begin{eqnarray}
\label{eq:AFieldTarget}
A^{-}(\xt,x^{+})=-\frac{\rho(\xt,x^{+})}{\nabt^2}\;,
\end{eqnarray}
where $\nabt^2=\partial_i\partial_i$ is the 2-dimensional Laplacian.
The scattering of an incoming quark inside the projectile can be described to leading order accuracy in $\alpha_s$ by the solution of the Dirac equation
\begin{eqnarray}
(i\slashed{D}-m) \hat{\Psi}=0\;,
\end{eqnarray}
in the presence of the background field of the target in Eq.~(\ref{eq:AFieldTarget}). One finds that the forward scattering amplitude of a quark with momentum $\pt$ to scatter off the color fields in the target is given by \footnote{Since we are primarily interested in the transverse coordinate dependence, we have omitted a delta function for longitudinal momentum conservation as well as the spin structure to lighten the notation. We refer to \cite{Dumitru:2002qt} for the complete expression.}
\begin{eqnarray}
\label{eq:forwardamplitude}
\left<{\rm out},\qt |{\rm in},\pt\right>= \int \ud^2\xt~V(\xt)~e^{i(\qt-\pt)\cdot\xt}\,,
\end{eqnarray}
where 
\begin{eqnarray}\label{eq:wilsonLine}
V(\xt)=\mathcal{P} \exp\left\{ -ig \int_{-\infty}^{\infty} \ud x^{+} A^{-}(x^{+},\xt) \right\}
\end{eqnarray}
denotes the Wilson line at a spatial position $\xt$  in the fundamental representation. 

Within the leading order dilute-dense framework, it is then straightforward to compute the single inclusive distributions of quarks in a high energy projectile after scattering from a nuclear target 
\begin{equation}
\frac{\ud N_{q}}{\ud^2\pt} = \left<{\rm out},\pt| \hat{\rho} | {\rm out},\pt \right> \;,
\end{equation}
where $ \hat{\rho}$ is the reduced one particle density matrix in the probe. This general expression can be rewritten explicitly as\footnote{Our expression generalizes the one given in \cite{Dumitru:2002qt} by replacing the collinear quark distribution with a Wigner function $W_{q}(\bt,\kt)$ that is a function of both the $\kt$ of quarks in the projectile and their impact parameter $\bt$. Equivalent expression for gluons, differing only by the representation of the Wilson lines, have explicitly been derived in \cite{Kharzeev:2003wz,Blaizot:2004wu}.}
\begin{multline}
\label{eq:SingleInclusive}
\frac{\ud N_{q}}{\ud^2\pt}= \int \ud^2\bt \int \frac{d^2\kt}{(2\pi)^2} \, W_{q}(\bt,\kt)  \varphi(\bt,\pt-\kt)\,.
\end{multline}
The Wigner function $W_{q}(\bt,\kt) $ characterizes the transverse momentum and position distribution of  incoming quarks inside the projectile and is defined to be 
\begin{equation}
W_{q}(\bt,\kt) = 
\int \frac{\ud^2\qt}{(2\pi)^2} 
\left<{\rm in},\kt+\frac{\qt}{2}\right| \hat{\rho} \left| {\rm in},\kt-\frac{\qt}{2} \right>
 e^{-i\qt\cdot\bt}\;.
\end{equation}
For the illustrative purpose of this paper, it is sufficient to choose a Gaussian form
\begin{equation}
W_{q}(\bt,\kt) = \frac{1}{\pi^2}~e^{-\bt^2/B}e^{-\kt^2 B} \, ,
\end{equation}
with a dimensionful constant $B$ characterizing the transverse area of the projectile. The dynamics of interest to us is given by the unintegrated gluon distribution of the target nucleus
\begin{equation}
\varphi(\bt,\kt)=\int \ud^2\rt~D \left( \bt+\frac{\rt}{2},\bt-\frac{\rt}{2}\right) ~e^{i\kt\cdot \rt}\;,
\end{equation}
which represents the distribution of momentum transfers from the target that contribute to the corresponding momentum distribution of the scattered quark. Here
\begin{align}
\label{eq:dipole}
D(\xt,\yt) &= \left\langle \mathcal{D} (\xt,\yt)\right\rangle
\end{align}
is the expectation value of the dipole operator
\begin{align}
\mathcal{D} (\xt,\yt) &= \frac{1}{\nc}  \tr \left[ V\left(\xt\right)
 V^{\dagger}\left(\yt\right) \right] \,,
\end{align}
 which results from the product of the forward scattering amplitude in Eq.~(\ref{eq:forwardamplitude}) and its complex conjugate equivalent, to obtain the single inclusive probability~\cite{McLerran:1998nk,Kovchegov:2001sc,Kovner:2001vi}.
Written fully in coordinate space our expression for the single inclusive
multiplicity~\nr{eq:SingleInclusive} has the form
\begin{multline}
\frac{\ud N_{q}}{\ud^2\pt}= \frac{1}{\pi B} \frac{1}{(2\pi)^2}
\int \ud^2\xt \ud^2\yt 
D(\xt,\yt) 
\\
e^{i \pt\cdot(\xt-\yt)}
\;  e^{-\frac{\xt^2}{2B}} 
\; e^{-\frac{\yt^2}{2B}},
\end{multline}
which is recognizable as the one used in \cite{Lappi:2015vha} up to a
constant factor which cancels in the anisotropy coefficients $v_n$.

Since we are interested in the scattering of a small probe -- such as a quark inside a proton -- off a large nucleus, it is reasonable to neglect the impact parameter $\bt=(\xt+\yt)/2$ dependence in the target.
Because on average there is no preferred direction in the transverse plane, the expectation value $D(\xt,\yt)$ then depends only on the magnitude of the transverse separation $\rt=\xt-\yt$ of the dipole\footnote{The dipole expectation value can in general depend on the rotational invariants $|\bt|$,$|\rt|$ and $\bt\cdot \rt$.  Once one performs the averaging over the $\bt$ distribution of the projectile, only the dependence on $\rt$ remains. We note however, that such an averaging can effectively introduce non-Gaussian correlations between several dipoles. These can affect multiparticle correlations at a characteristic momentum scale given by the inverse impact parameter dependence.}, 
\begin{eqnarray}
\label{eq:AvgDipole}
D(\xt,\yt)  \equiv D(|\xt-\yt|) \;.
\end{eqnarray} 
Since the dipole coordinate $\rt$ is the conjugate variable to the momentum transfer $\pt-\kt$ for a single quark scattering, the symmetry in Eq.~(\ref{eq:AvgDipole}) ensures that the momentum transfer to each individual quark is on average symmetric with respect to the azimuthal angle.\\ 

\subsection{Double inclusive spectrum and multiparticle correlations}
We will now consider the case where two quarks scatter independently off the same nucleus and shall study the correlations between the two scattered quarks. We will make the simplifying assumption that the momenta of the two incoming quarks are initially uncorrelated~\footnote{We are mostly concentrating on the near-side ``ridge'' correlation for semihard momenta $\sim \qs$. Thus we are neglecting back-to-back momentum correlations~\cite{JalilianMarian:2004da,Baier:2005dv} that are particularly important for the away-side ``jet'' peak~\cite{Albacete:2010pg,Dusling:2012cg,Lappi:2012nh} and contributions at the nonperturbative small intrinsic transverse  momentum scale of the probe. Such correlations should be taken into account in a full comparison with data.} - the two particle distribution $ W_{qq}(\bt_1,\kt_1,\bt_2,\kt_2)$ of incoming quarks factorizes into the product of single quark distributions, 
\begin{eqnarray}
W_{qq}(\bt_1,\kt_1,\bt_2,\kt_2)=W_{q}(\bt_1,\kt_1)W_{q}(\bt_2,\kt_2)\;. \nonumber \\
\end{eqnarray}

The double inclusive distribution of scattered quarks then takes the form
\begin{multline}
\label{eq:DoubleInclusive}
\frac{\ud^2N}{\ud^2\pt_1 \ud^2\pt_2}=  \int \ud^2\bt_1 \ud^2\bt_2 \int \frac{\ud^2\kt_1}{(2\pi)^2} \int \frac{\ud^2\kt_2}{(2\pi)^2}  
 \\
\times
 W_{q}(\bt_1,\kt_1)  W_{q}(\bt_2,\kt_2)  
 \\
\times \int \ud^2\rt_1 \ud^2\rt_2  
e^{i(\pt_1-\kt_1) \cdot \rt_1} e^{i(\pt_2-\kt_2)\cdot \rt_2}
 \\
\times
\bigg< \mathcal{D}\left(\bt_1+\frac{\rt_1}{2},\bt_1-\frac{\rt_1}{2}\right)
\\
\times \mathcal{D}\left(\bt_2+ \frac{\rt_2}{2},\bt_2- \frac{\rt_2}{2}\right)
\bigg> .
\end{multline}
While we assumed the transverse momenta $\kt_1$ and $\kt_2$ of the two incoming quarks to be uncorrelated, one observes from Eq.~(\ref{eq:DoubleInclusive}) that this is no longer the case for the momenta $\pt_1$ and $\pt_2$ of the scattered quarks. Since both quarks scatter off the same nucleus, the momentum transfers $\pt_1-\kt_1$ and $\pt_2-\kt_2$ are correlated with each other, giving rise to azimuthal correlations of the scattered quarks.  \\


We note that the above model can be generalized in a straightforward way to study correlation functions involving more than two particles. Such higher order correlation functions can be related to higher order correlation functions of dipole correlators. For example, four quark correlations in this model involve expectation values of products of four dipoles in the fundamental representation.

\section{Dipole-dipole correlator}
The discussion in the previous section shows that all the features of two-particle
correlations are encoded in the expectation value of the dipole-dipole correlator $\langle \mathcal{D}(\xt,\yt) \mathcal{D}(\ut,\vt)\rangle$. We will now study the properties of this correlator and discuss approximation schemes that have been frequently employed in the literature. 

\subsection{Glasma graph approximation}

The basic properties of the two particle correlation can be understood by studying the interaction between the incoming quark and the target nucleus in terms of  multiple gluon exchanges. Clearly the dynamics of these gluon exchanges depends on the nature of color charge correlations in the target and needs to be specified.  A simple model of such correlations is the MV model~\cite{McLerran:1994ni,McLerran:1994ka,McLerran:1994vd}. In this model, the underlying distribution of color charges in the nucleus is assumed to be a Gaussian distribution such that  all multigluon correlations are uniquely determined by the two gluon correlation function.

A further approximation that simplifies the computation considerably is to assume that each of the quarks in the projectile exchanges only two gluons with the target nucleus. We will refer to this combination of the two gluon exchange approximation and Gaussian statistics as the Glasma graph approximation, a term first introduced in Ref.~\cite{Dusling:2012iga}. This approximation has been used in a number of phenomenological studies of ridge correlations in high energy collisions~\cite{Dumitru:2008wn,Gavin:2008ev,Dusling:2009ar,Gelis:2009wh,Dusling:2009ni,Dumitru:2010iy,Dusling:2012iga,Dusling:2012wy,Dusling:2013oia}.

More specifically, the Glasma graph approximation can be understood by examining the expression for the dipole-dipole correlator in terms of the color field. When the density of color charges in the target nucleus $g\rho$ is small -- corresponding to a dilute-dilute situation -- one can perform an expansion of the path ordered Wilson line around the identity matrix, representing an expansion in terms of the number of gluons exchanged between the projectile and the target. In order to keep the notation as light as possible, we will denote the Wilson lines as $V(\xt)=\exp(-i\Lambda(\xt))$  in the following\footnote{Careful path ordering will not modify the results obtained as long as the correlations are local in $x^+$.}, so that the expansion takes the form 
\begin{eqnarray}
V(\xt)&\simeq& \id -i \Lambda(\xt) -\frac{1}{2}  \Lambda^{2}(\xt) \\
&&+\frac{i}{6} \Lambda^{3}(\xt)+\frac{1}{24} \Lambda^{4}(\xt) + \dots \;. \nonumber
\end{eqnarray}
Evaluating the color traces for $\Lambda(\xt) = \Lambda^a(\xt) t^a$ as
\begin{eqnarray}
\text{tr}[t^{a}t^{b}] = \frac{\delta^{ab}}{2}\;, \quad  \text{tr}[t^{a}t^{b}t^{c}] =\frac{1}{4}(if^{abc}+d^{abc})\;,
\end{eqnarray}
the dipole operator  to $\mathcal{O}(\Lambda^3)$ within this approximation takes the form
\begin{multline}
\label{eq:dipoleexpansion}
\mathcal{D}(\xt,\yt)-1\simeq 
\\
 - \frac{1}{2} \frac{\delta^{ab}}{2\nc}  \Big(\Lambda^{a}_{\xt}\Lambda^{b}_{\xt}- 2\Lambda^{a}_{\xt}\Lambda^{b}_{\yt} +\Lambda^{a}_{\yt}\Lambda^{b}_{\yt}\Big)
 \\
 +\frac{i}{6} \frac{d^{abc}}{4\nc} \Big(\Lambda^{a}_{\xt}\Lambda^{b}_{\xt}\Lambda^{c}_{\xt} -3\Lambda^{a}_{\xt}\Lambda^{b}_{\xt}\Lambda^{c}_{\yt} 
 \\
 + 3\Lambda^{a}_{\xt}\Lambda^{b}_{\yt}\Lambda^{c}_{\yt} - 
\Lambda^{a}_{\yt}\Lambda^{b}_{\yt}\Lambda^{c}_{\yt} \Big) + \dots \;.   
\end{multline}
In the Gaussian approximation for the correlations of $\Lambda$ the
$\mathcal{C}$ and $\mathcal{P}$ odd $\mathcal{O}(\Lambda^3)$ term vanishes in the expectation value 
of the dipole operator and only the $\mathcal{O}(\Lambda^2)$ remains. We will denote the $\Lambda\Lambda$ correlator as 
\begin{equation}
\langle \Lambda_{a}(\xt) \Lambda_{b}(\yt) \rangle= \delta^{ab} \gamma(\xt-\yt)\;, 
\label{eq:defgamma}
 \end{equation}
which defines the correlation function $\gamma(\rt)$.
With this definition, we obtain~\cite{McLerran:1998nk}
\begin{equation}
\label{eq:dipole-average}
D(\xt-\yt)
\simeq 1- \cf 
\Big( \gamma(\ot)-\gamma(\xt-\yt) \Big)\;. \nonumber \\
\end{equation}

\begin{figure}[t!]
   \begin{center}   
     	
       \includegraphics[width=0.48\textwidth]{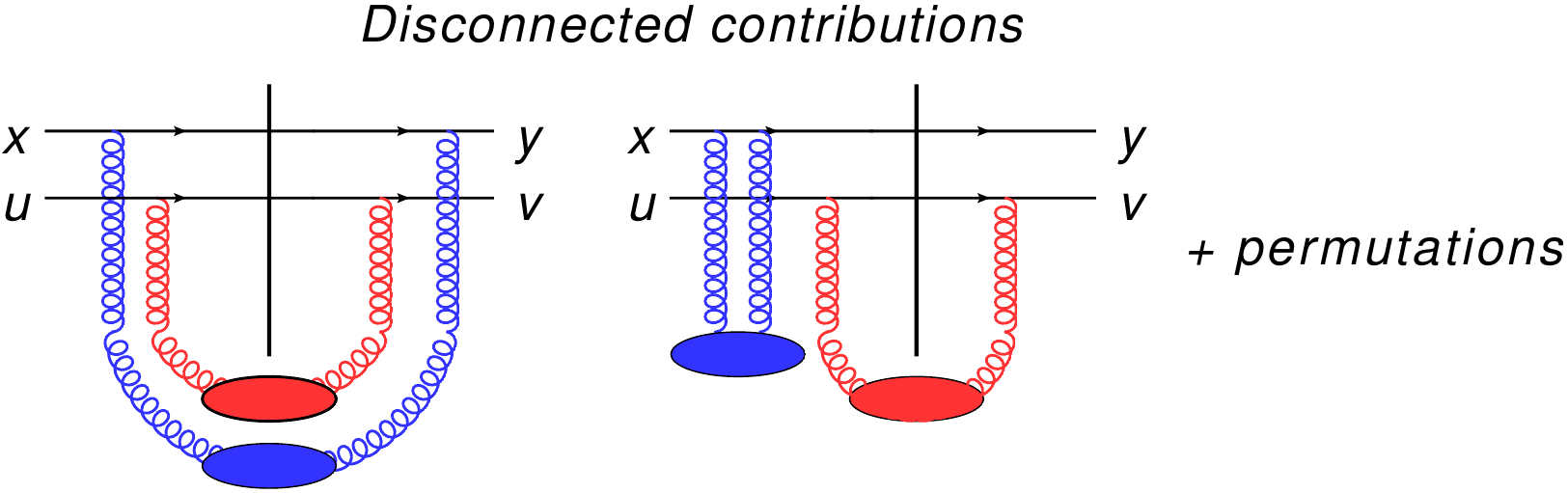} 
	 
        \includegraphics[width=0.4\textwidth]{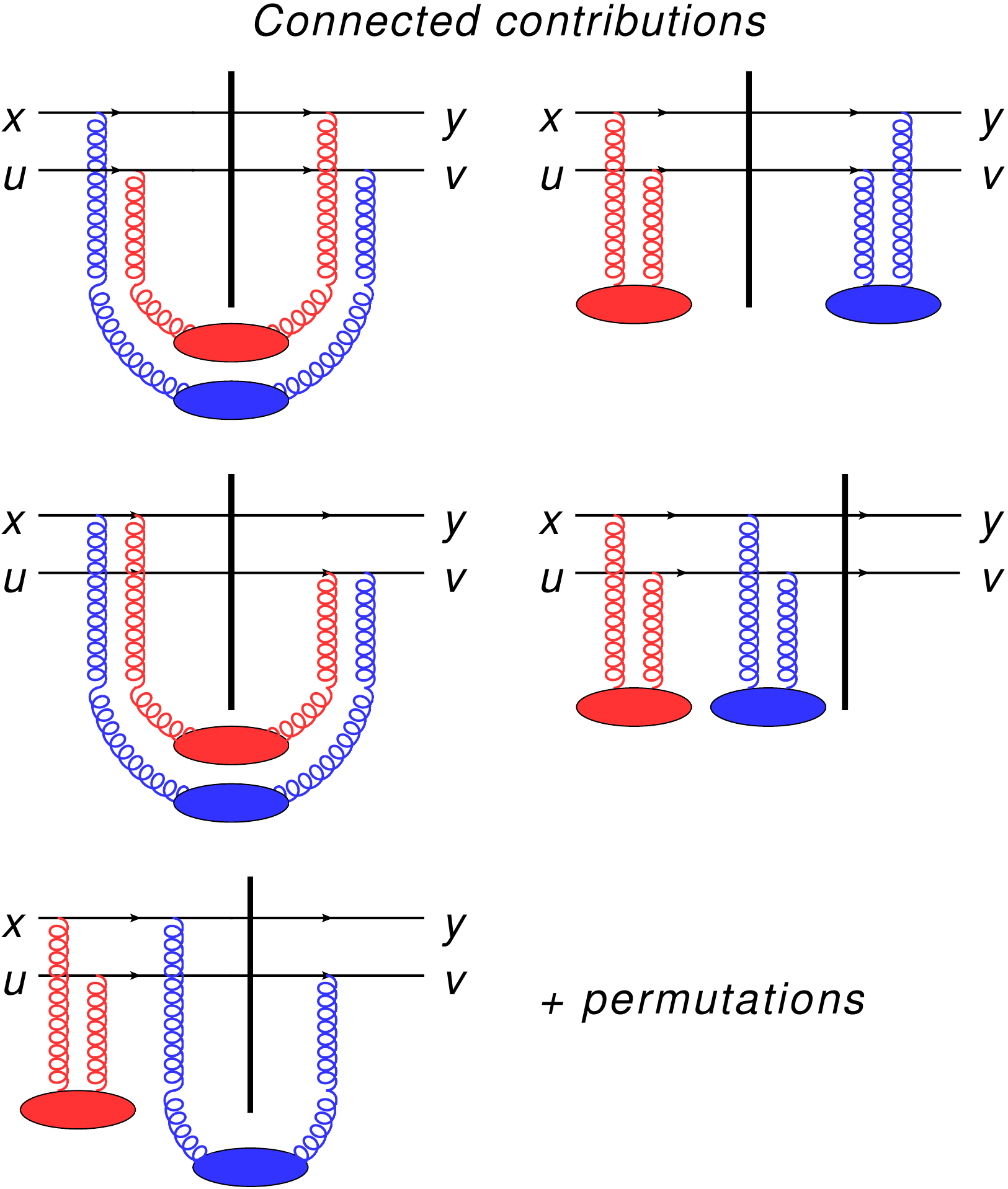}

     \caption{ \label{fig:Diagrams2g} (Color online) Classes of diagrams contributing to the dipole-dipole correlator in double-gluon exchange approximation.}
      \end{center}
\end{figure}

Following the same logic as previously for the dipole expectation value, we will now compute the dipole-dipole correlator by expanding up to $\mathcal{O}(\Lambda^4)$ in the coupling constant. The assumption of Gaussian statistics means that the four point correlation function can be expressed in terms of the two point function in Eq.~\nr{eq:defgamma} as
\begin{align}
& \langle \Lambda_{a}(\xt) \Lambda_{b}(\yt)  \Lambda_{c}(\ut) \Lambda_{d}(\vt) \rangle=  
\\ \nonumber
&\qquad\qquad \quad \delta^{ab}\delta^{cd} \gamma(\xt-\yt) \gamma(\ut-\vt) 
\\ \nonumber
&\qquad\qquad +\delta^{ac}\delta^{bd} \gamma(\xt-\ut) \gamma(\yt-\vt) 
 \\ \nonumber
&\qquad\qquad +\delta^{ad}\delta^{bc} \gamma(\xt-\vt) \gamma(\yt-\ut)\  .
\end{align}
Thus one obtains
\begin{multline}
\langle \mathcal{D}(\xt,\yt) \mathcal{D}(\ut,\vt) \rangle \simeq D(\xt-\yt) D(\ut-\vt)  \\
+\frac{\cf^2}{2 (\nc^2-1)} \Big( \gamma(\xt-\ut)-\gamma(\xt-\vt) 
\\
  -\gamma(\yt-\ut)+\gamma(\yt-\vt)\Big)^2\;,
\label{eq:ggwithgamma}
\end{multline}
which, to this order of the approximation, is consistent with
\begin{multline}
\label{eq:SingleGluonExchange}
\langle \mathcal{D}(\xt,\yt) \mathcal{D}(\ut,\vt) \rangle \simeq D(\xt-\yt) D(\ut-\vt) \\
 +\frac{1}{2(\nc^2-1)} \Big(D(\xt-\vt)-D(\xt-\ut)  \\
 -D(\yt-\vt)+D(\yt-\ut)\Big)^2\,.
\end{multline} 
One observes from Eq.~(\ref{eq:SingleGluonExchange}) that the leading $\nc$ contribution to the dipole-dipole correlator factorizes into the product of single inclusive averages -- corresponding to the independent scattering of two quarks. Diagrammatically this corresponds to the disconnected contribution in \fig\ref{fig:Diagrams2g}.  Genuine correlations are contained in the second term of Eq.~(\ref{eq:SingleGluonExchange}) and suppressed by a factor of $1/(\nc^2-1)$ as pointed out previously in \cite{Dumitru:2008wn,Dusling:2009ar,Gelis:2009wh}.  These correspond diagrammatically to the connected graphs in \fig\ref{fig:Diagrams2g} and feature all possible contractions between the $\xt,\yt$ and $\ut,\vt$ dipoles.

When calculating two particle correlations using Eq.~\nr{eq:DoubleInclusive}, one needs a Fourier transform of the dipole-dipole correlator in Eq.~\nr{eq:SingleGluonExchange} to obtain the momentum transfer $\pt-\kt$ from the interaction with the target. One finds that the terms in Eq.\nr{eq:SingleGluonExchange} that do not depend on all four coordinates $\xt,\yt,\ut,\vt$ only contribute for very soft momenta $\pt,\kt$ on the order of the incoming momentum of the projectile $\ptt,\qtt \sim 1/\sqrt{B}$. Since we are focusing on the dominant semihard gluons with $\ptt,\qtt \gtrsim \qs \gg  1/\sqrt{B}$ we will neglect these contributions in the following and replace \eq\nr{eq:SingleGluonExchange} with 
\begin{multline}
\label{eq:SingleGluonExchangev2}
\langle \mathcal{D}(\xt,\yt) \mathcal{D}(\ut,\vt) \rangle \simeq D(\xt-\yt) D(\ut-\vt) \\
 +\frac{1}{(\nc^2-1)} \Big(D(\yt-\ut)D(\xt-\vt)  \\
 +D(\yt-\vt)D(\xt-\ut) \Big)\,.
\end{multline} 
Equation \nr{eq:SingleGluonExchangev2} is in fact the form used in the Glasma graph 
papers~\cite{Dumitru:2008wn,Gavin:2008ev,Dusling:2009ar,Gelis:2009wh,Dusling:2009ni,Dumitru:2010iy,Dusling:2012iga,Dusling:2012wy,Dusling:2013oia}.
In particular, in the comparisons to data performed in \cite{Dusling:2012iga,Dusling:2012wy,Dusling:2013oia}, the expectation value of the dipole correlator is taken to satisfy the 
Balitsky-Kovchegov (BK) equation~\cite{Balitsky:1995ub,Kovchegov:1999yj}. 
Thus the Glasma graph approximation, as employed in phenomenological computations, corresponds to the merging of gluon ladders from multiple sources (all localized on a transverse scale $\sim 1/\qs$) into a single ladder exchange represented by the dipole correlator. 

\subsection{Nonlinear Gaussian approximation}
If we restrict ourselves to Gaussian correlations of gluon fields in the target, it is possible to resum the multiple gluon exchange contributions to the dipole amplitude and the dipole-dipole correlator analytically to all orders\footnote{As suggested by our previous discussion, these become multiple \emph{ladder} exchanges upon BK or JIMWLK evolution of the dipole and dipole-dipole correlators.}. Generalizing the notation from the glasma graph case, one obtains~\cite{McLerran:1998nk}
\begin{equation}\label{eq:nonlindipole}
D(\xt-\yt)=\exp\Big(\cf \left(\gamma(\xt-\yt)-\gamma(\ot)\right) \Big) \, .
\end{equation}
This expression is a part of the \emph{nonlinear} Gaussian approximation because it 
includes all orders in $\Lambda$, evaluated
with a the Gaussian $\Lambda\Lambda$ correlator in \eq\nr{eq:defgamma}. 
This is to be contrasted  with the two-gluon exchange
approximation  \eq\nr{eq:dipole-average} which was expanded to the lowest order.
Using a well known algorithm~\cite{Kovner:2001vi,Blaizot:2004wu,Blaizot:2004wv,Kovchegov:2008mk,Fujii:2006ab,Marquet:2010cf,Dominguez:2011wm,Iancu:2011ns,Iancu:2011nj,Dominguez:2011wm}
 for computing higher point correlators one can obtain the dipole-dipole operator expectation value to all orders in $\Lambda$ assuming a Gaussian  $\Lambda\Lambda$ correlator as~\cite{Dominguez:2008aa} 
\begin{align}
\label{eq:NonLinearGaussian}
\nonumber
&
\langle \mathcal{D}(\xt,\yt) \mathcal{D}(\ut,\vt) \rangle= 
D(\xt-\yt) D(\ut-\vt) 
\\ \nonumber
&
\quad 
\times  \Bigg[ \left( \frac{ F{(\xt,\ut;\yt,\vt)}+\sqrt{\Delta}}{2 \sqrt{\Delta}} 
- \frac{F{(\xt,\yt;\ut,\vt)}}{\nc^2 \sqrt{\Delta}} \right) 
e^{\frac{\nc}{4} \sqrt{\Delta}} 
\\ \nonumber
&
 \quad \quad 
-\left( \frac{F{(\xt,\ut;\yt,\vt)}-\sqrt{\Delta}}{2\sqrt{\Delta}} 
- \frac{F{(\xt,\yt;\ut,\vt)}}{\nc^2 \sqrt{\Delta}} \right) 
 e^{-\frac{\nc}{4} \sqrt{\Delta}} \Bigg] 
 \\ 
&
\quad \quad \quad \quad \quad \quad 
 \times e^{-\frac{\nc}{4} F{(\xt,\ut;\yt,\vt)}+ \frac{1}{2\nc} F{(\xt,\yt;\ut,\vt)}}\;,
\end{align}
where $F{(\xt,\yt;\ut,\vt)}$ is defined to be 
\begin{equation}
F{(\xt,\yt;\ut,\vt)}= 
\frac{1}{\cf}\ln \left( \frac{D(\xt-\ut) D(\yt-\vt)}{D(\xt-\vt)D(\yt-\ut)} \right)\;. 
\end{equation}
Here $\Delta$ is short for $\Delta(\xt,\yt;\ut,\vt)$ and is given by
\begin{multline}
\Delta(\xt,\yt;\ut,\vt)=F^{2}{(\xt,\ut;\yt,\vt)}  \\
 +\frac{4}{\nc^2} F{(\xt,\yt;\ut,\vt)}F{(\xt,\vt;\ut,\yt)}\;.
\end{multline}
We will refer to Eq.~(\ref{eq:NonLinearGaussian}) as the ``nonlinear Gaussian 
approximation'' for the dipole-dipole correlator. It is exact in the 
McLerran-Venugopalan (MV) model where
dipole and multipole correlators depend nonlinearly on the $\Lambda$ fields, but 
the correlators of $\Lambda$'s are Gaussian.
One can easily check that the two-gluon exchange limit of the nonlinear 
Gaussian~\nr{eq:NonLinearGaussian} is the same
as the two-gluon exchange approximation \nr{eq:ggwithgamma}.

One can obtain sight into this relation by taking the large $\nc$ limit for a constant $D(\xt-\yt)$:
\begin{multline}
\langle \mathcal{D}(\xt,\yt) \mathcal{D}(\ut,\vt) \rangle = D(\xt-\yt)D(\ut-\vt)  
\\
+\frac{1}{\nc^2} 
 \left( \frac{   \ln \left(   \frac{D(\xt-\ut)D(\yt-\vt)}{D(\xt-\vt)D(\ut-\yt)} \right)                   }   { \ln \left(   \frac{D(\xt-\yt)D(\ut-\vt)}{D(\xt-\vt)D(\ut-\yt)} \right)   } \right)^2
\Bigg[ D(\xt-\vt) D(\ut-\yt)  
\\
+D(\xt-\yt)D(\ut-\vt) 
 \left( \ln \left(   \frac{D(\xt-\yt)D(\ut-\vt)}{D(\xt-\vt)D(\ut-\yt)} \right)  -1 \right) \Bigg]    
\;,
\label{eq:nonlingausslargenc}
\end{multline}
which shows again that the leading $\nc$ contribution corresponds to the independent scattering of two quarks, while genuine correlations are $\nc$ suppressed.

\begin{figure}[t!]
   \begin{center}   
     	
       \includegraphics[width=0.48\textwidth]{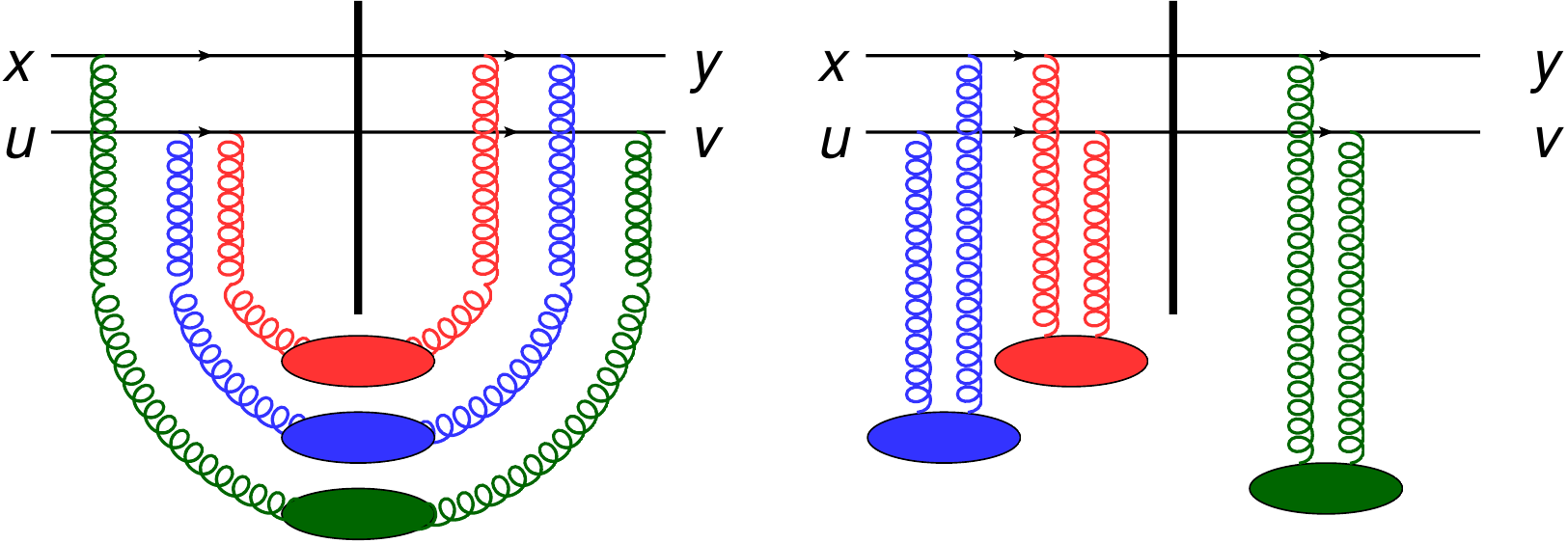} 

     \caption{ \label{fig:DiagramsAS} (Color online) Example of a higher order  contribution to the non-linear Gaussian approximation, which contributes to the $\xt\leftrightarrow\yt$ anti-symmetric part of the dipole-dipole correlator at $\mathcal{O}(\Lambda^6)$. }
      \end{center}
\end{figure}

While the nonlinear Gaussian approximation reproduces the double gluon exchange (Glasma graph) approximation at $\mathcal{O}(\Lambda^4)$ in the dilute limit, it also contains a series of higher order terms. An important subset of higher order contributions corresponds to the diagrams which separately break the $\xt-\yt \to \yt-\xt$ and $\ut-\vt \to \vt-\ut$ symmetries. As we will discuss shortly these contributions are responsible for generating the odd moments $(v_3,v_5,...)$ in the Fourier expansion of the correlation function. One finds that the leading contribution to the $\xt\leftrightarrow \yt$ antisymmetric part can be associated with the square of the $\mathcal{C}$ and $\mathcal{P}$ odd contribution to the dipole operator in \eq\nr{eq:dipoleexpansion}.
Note that while the expectation value of the odd term is zero in the Gaussian
approximation, the expectation value of its square is not, but is proportional
to $d^{abc}d^{abc} = (\nc^2-1)(\nc^2-4)/\nc$.
 The contributions from the odd terms correspond diagrammatically to the processes depicted in Fig.\,\ref{fig:DiagramsAS} and take the form
\begin{eqnarray}
\langle \mathcal{D}(\xt,\yt) \mathcal{D}(\ut,\vt) \rangle - \langle \mathcal{D}(\yt,\xt) \mathcal{D}(\ut,\vt) \rangle \simeq \nonumber \\
 -\frac{\cf^3}{6} \frac{\nc^2-4}{(\nc^2-1)^2}\Big( \gamma(\xt-\ut)-\gamma(\yt-\ut) \nonumber \\
 -\gamma(\xt-\vt)+\gamma(\yt-\vt)\Big)^3\;.
\end{eqnarray}
While for $\nc=2$ the antisymmetric contribution vanishes identically to all orders, it is nonvanishing for $\nc\geq3$ and suppressed by a factor of $1/\nc^2$ relative to the disconnected contribution in the large $\nc$ limit.

\section{Azimuthal correlations in quark nucleus scattering}

We will now discuss the azimuthal correlations of quarks scattering off a large nucleus and  present comparisons of results for these correlations within different approximation schemes introduced in the previous sections. To further quantify the correlations introduced by the scattering,  we will decompose 
the double inclusive distribution in  Eq.~(\ref{eq:DoubleInclusive}) into Fourier modes in the relative azimuthal angle $\Delta \phi$ between the two scattered quarks, 
\begin{equation}\label{eq:FourierDec}
\frac{\ud^2N}{\ud^2\pt_1 \ud^2\pt_2} \propto 1 + \sum_{n=1}^\infty 2\, V_{n\Delta}(\pt_1,\pt_2)\cos(n\Delta\phi)\,.
\end{equation}
The familiar coefficients $v_n\{2\}(\ptt)$ can be obtained from the $V_{n\Delta}$ using \cite{Chatrchyan:2013nka}
\begin{equation}\label{eq:v22}
  v_n\{2\}(\ptt) = \frac{V_{n\Delta}(\ptt,\ptt^{\rm Ref})}{\sqrt{V_{n\Delta}(\ptt^{\rm Ref},\ptt^{\rm Ref})}}\,,
\end{equation}
where $p_T^{\rm Ref}$ represents a range in transverse momentum corresponding to an experimental reference bin. 

\subsection{Analytic estimates}
Before we turn to a discussion of numerical results, it is useful to obtain further analytic insight into the correlation functions themselves by considering the limit of a large probe with small intrinsic transverse momentum ($\qs^2 B \gg 1$). With the Gaussian Wigner distribution introduced in Sec.~\ref{sec:hybridcorr} it is convenient to absorb a part of the Wigner distribution into 
the definition of a modified dipole distribution
\begin{eqnarray}
\tilde{\gamma}(\pt)=\int \ud^2\rt~e^{-\frac{\rt^2}{4B}} 
D(r)~e^{i\pt\cdot\rt}\,.
\end{eqnarray}
such that the single inclusive distribution~\nr{eq:SingleInclusive} becomes
\begin{equation}
 \frac{\ud N_{q}}{\ud^2\pt}= \frac{1}{(2\pi)^2}\tilde{\gamma}(\pt)\;.
\end{equation}
Within the Glasma graph approximation the double inclusive distribution can then be evaluated by combining \eqs\nr{eq:DoubleInclusive} and~\nr{eq:SingleGluonExchangev2} as
\begin{multline}\label{eq:GGA}
 \frac{\ud^2N}{\ud^2\pt \ud^2\qt}=  
\frac{1}{(2 \pi)^4} \Bigg\{ \tilde{\gamma}(\pt) \tilde{\gamma}(\qt)
\\
+ 
\frac{1}{(\nc^2-1)}
\Bigg[
e^{-(\pt+\qt)^2B/2} \; 
\left(\tilde{\gamma} \left( \frac{\pt-\qt}{2} \right) \right)^2
\\
+
e^{-(\pt-\qt)^2B/2}  \; 
\left(\tilde{\gamma}\left( \frac{\pt+\qt}{2}\right)
\right)^2\Bigg]
\Bigg\}\;,
\end{multline}
which further reduces to
\begin{multline}
\label{eq:GGAcoll}
\frac{\ud^2 N}{\ud^2\pt \ud^2\qt}= 
\frac{1}{(2 \pi)^4}\tilde{\gamma}(\pt) \tilde{\gamma}(\qt)  
\\
\times \left[1+ \frac{2\pi  \left( \delta^{(2)}(\pt+\qt) + \delta^{(2)}(\pt-\qt) \right)  }{B(\nc^2-1)}\right]\,,
\end{multline}
in the limit of a large probe or at high momenta, where the intrinsic transverse momentum of the probe can be neglected and we can approximate
$2\pi B e^{-\frac{B}{2} \kt^2} \to (2\pi)^2 \delta^{(2)}(\kt) $.
While the first term inside the square bracket corresponds to the disconnected contribution and does not contain any correlations, the connected term gives rise to a correlation which is suppressed by $1/(\nc^2-1)$ and $1/(\qs^2B)$. 

One also observes from \eqs\nr{eq:GGA} and~\nr{eq:GGAcoll} that the two particle correlation function has a  similar structure as the collinear limit of the Glasma graph computation~\cite{Dumitru:2008wn,Gavin:2008ev,Dusling:2009ar,Gelis:2009wh,Dusling:2009ni,Dumitru:2010iy,Dusling:2012iga,Dusling:2012wy,Dusling:2013oia}. It features a near side ($\pt \approx \qt$) contribution as well as one on the away side  ($\pt \approx -\qt$), which in terms of the Fourier decomposition in Eq.~(\ref{eq:FourierDec}) give rise to even harmonics $v_{2},v_{4},...$. Odd harmonics $v_{3},v_{5},...$ are not present in \eqs\nr{eq:GGA} and~\nr{eq:GGAcoll} as the above expressions are manifestly symmetric under $\pt\to-\pt$.

\subsection{Numerical results}

\begin{figure*}[t!]   	 
     	\includegraphics[width=0.45\textwidth]{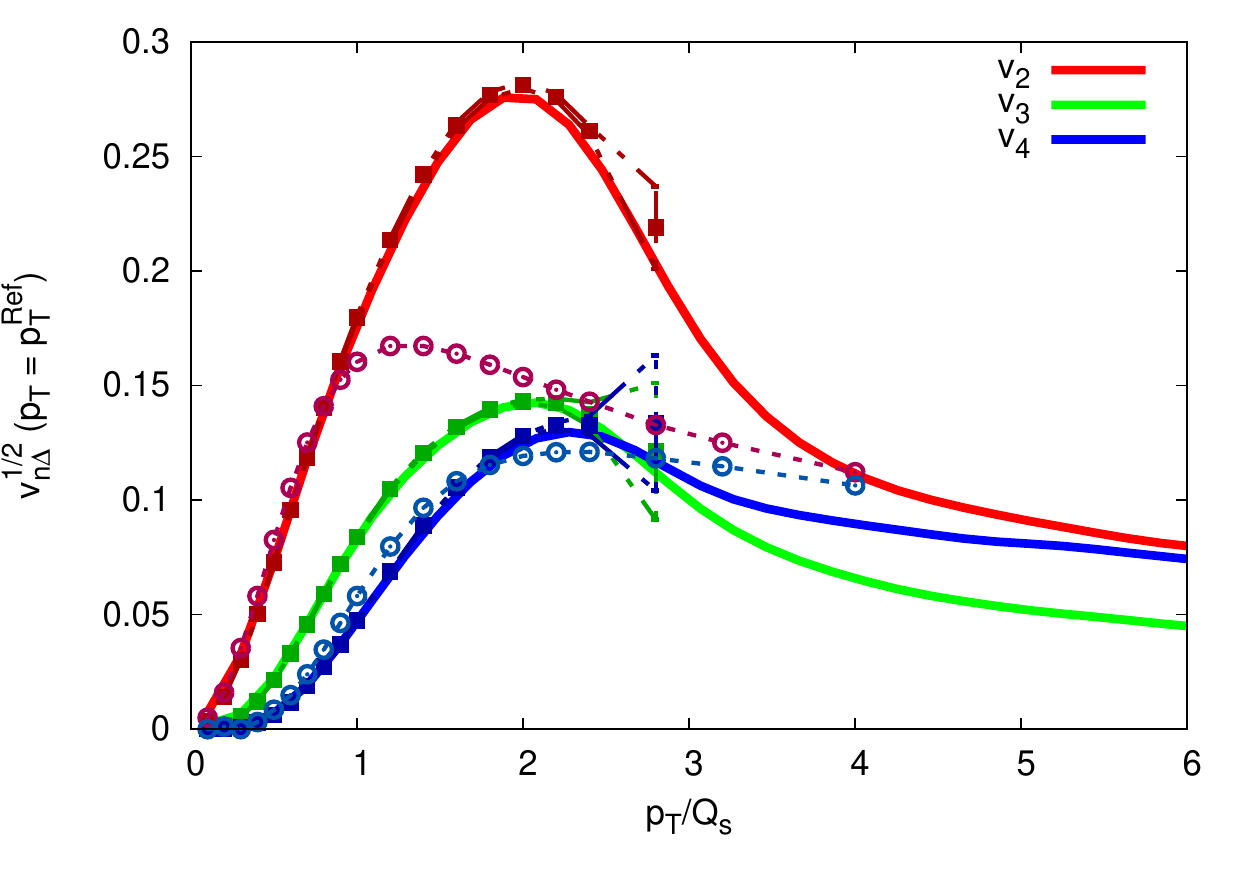} 	 
     	\includegraphics[width=0.45\textwidth]{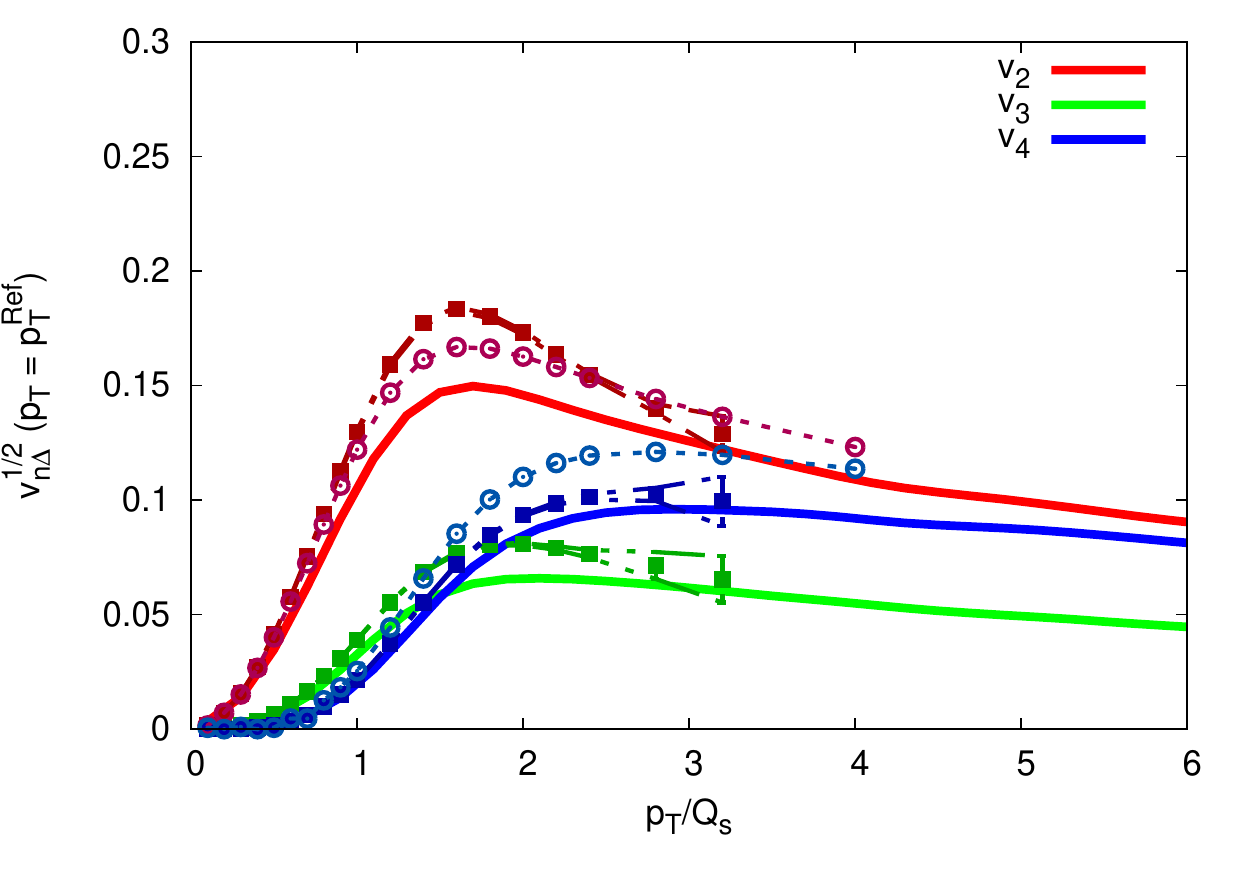} 

     \caption{ \label{fig:vNModel} (Color online) Azimuthal correlations $v_{n}$ for scattering of two independent quarks off a large nucleus in the MV model (left) and after JIMWLK evolution (right).  Solid lines correspond to results from the lattice simulation without additional approximation, dash-dotted lines (with squares) show results in the non-linear Gaussian approximation, and dotted lines (with circles) correspond to the Glasma graph approximation.}
\end{figure*}

\begin{figure*}[htb]   	 
     	\includegraphics[width=0.45\textwidth]{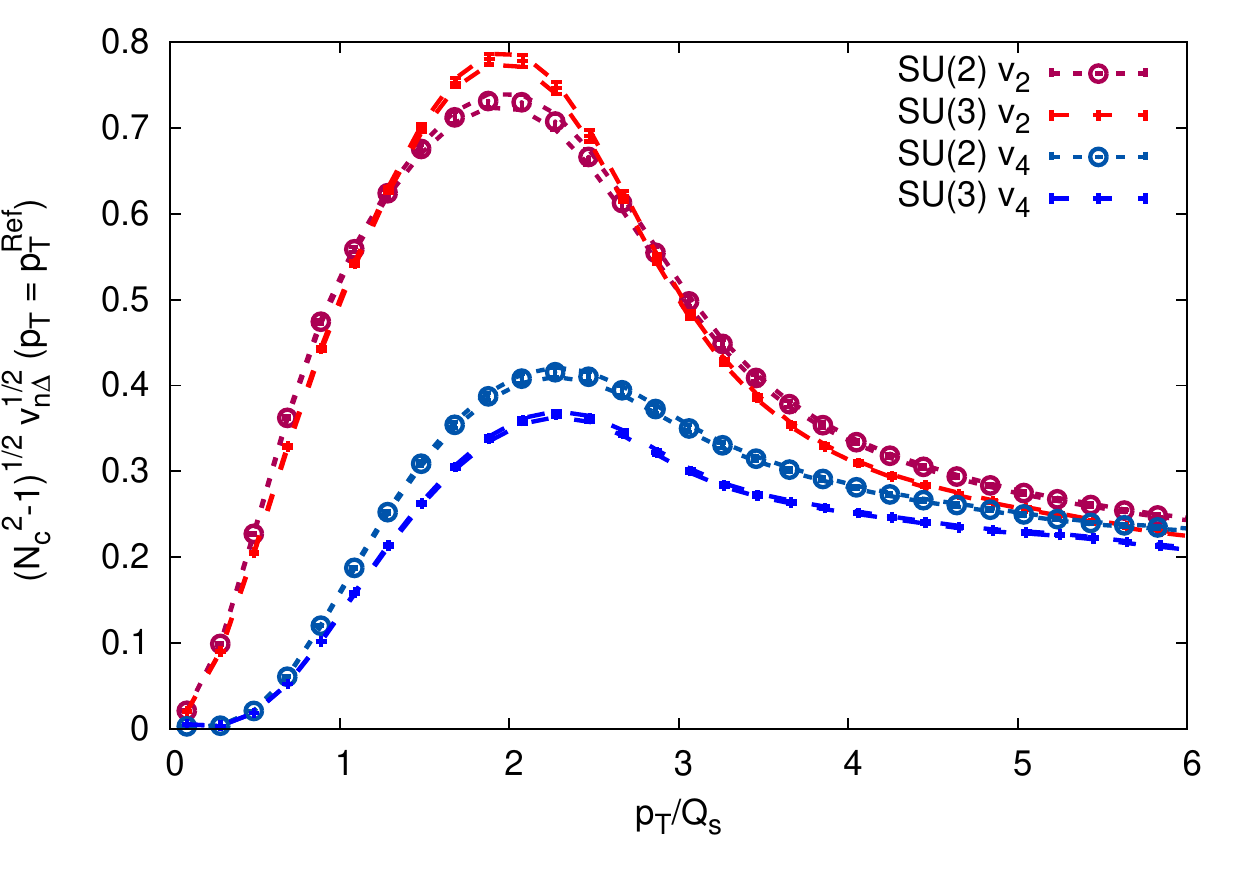} 	 
     	\includegraphics[width=0.45\textwidth]{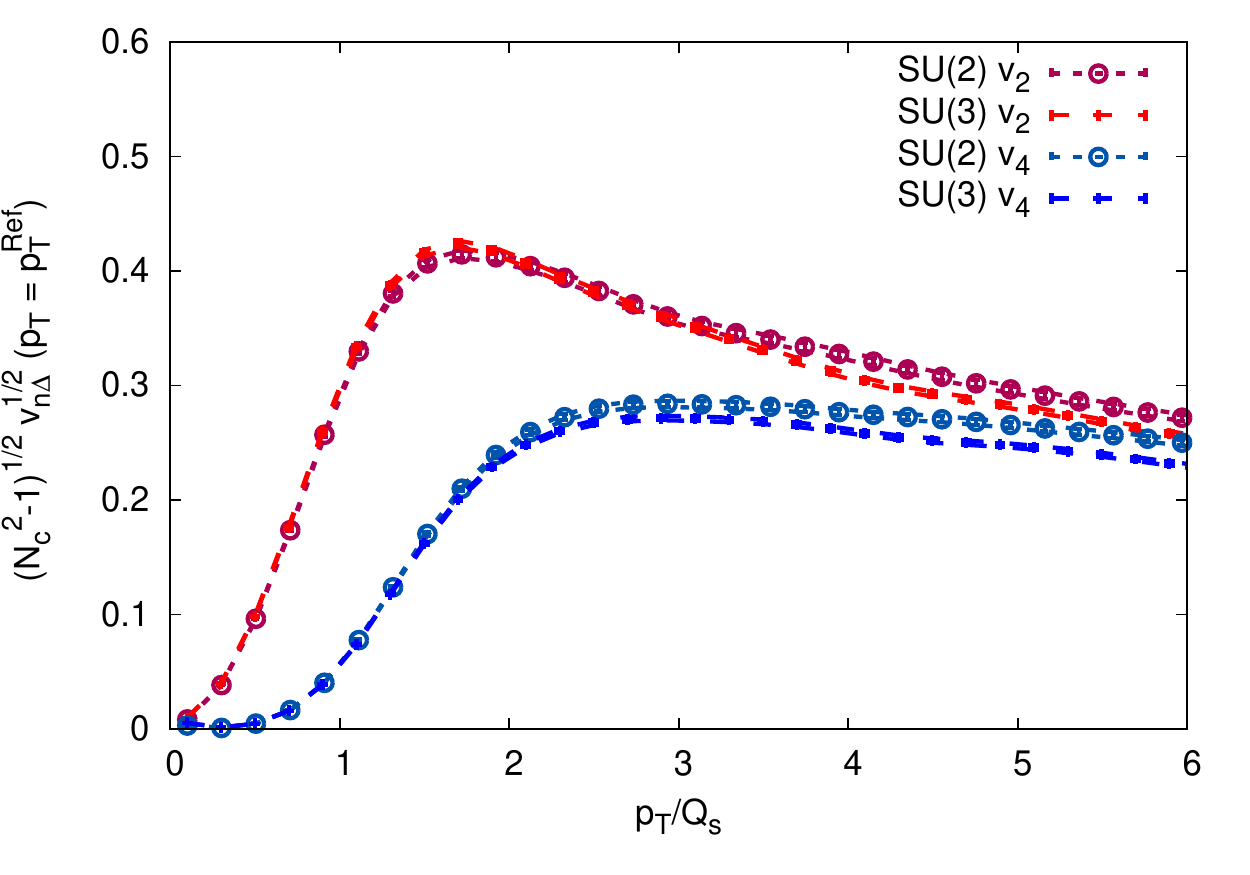} 

     \caption{ \label{fig:vnNc} (Color online) $\nc$ dependence of the azimuthal correlations $v_{n}$, scaled by the color factor $\sqrt{\nc^2-1}$ for SU(2) and SU(3)  gauge theory. The results are computed using the numerical lattice calculation for the MV model (left) and after JIMWLK rapidity evolution (right).}
\end{figure*}

\begin{figure*}[htb]   	 
     	\includegraphics[width=0.45\textwidth]{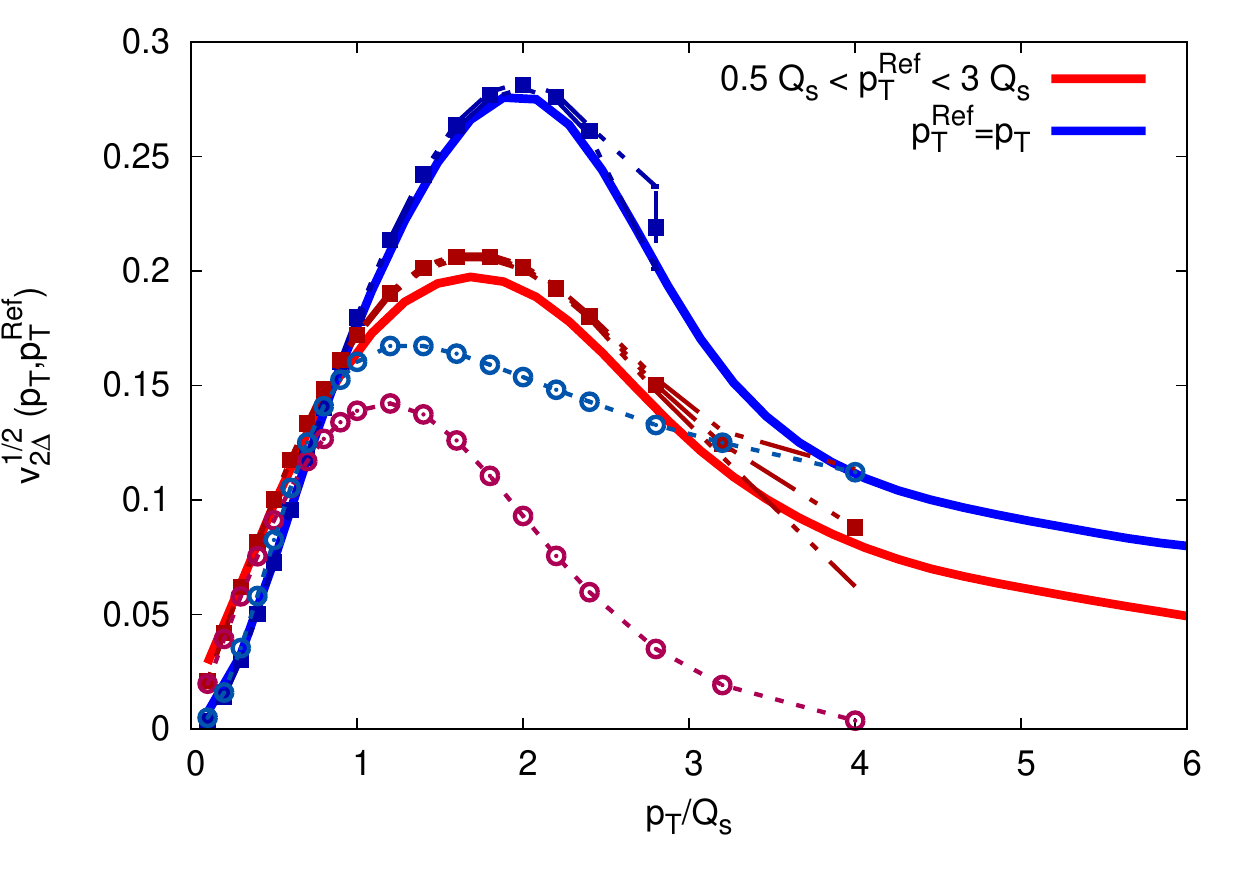} 	 
     	\includegraphics[width=0.45\textwidth]{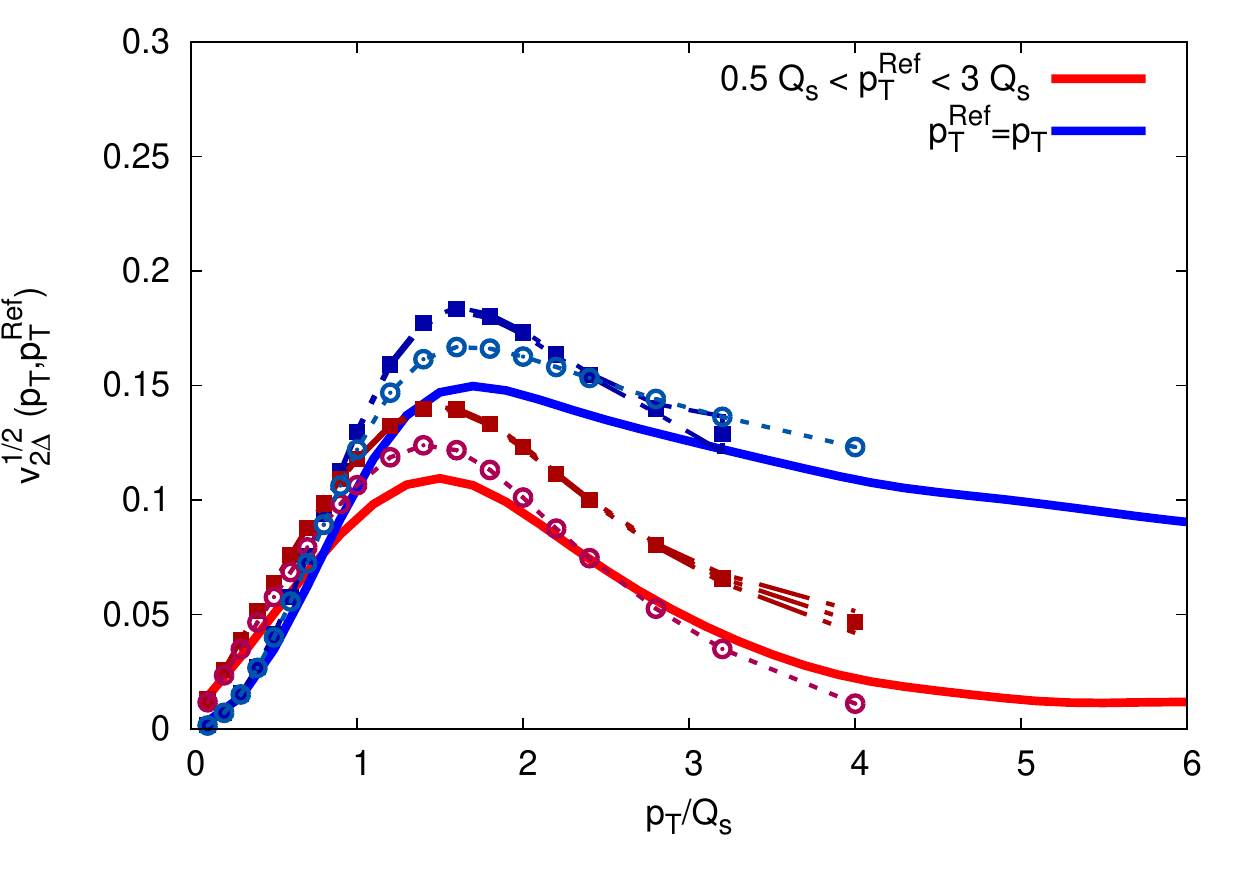} 

     \caption{ \label{fig:vnpT} (Color online) Comparison of azimuthal correlations $v_{2}$ in three different approximations for different choices of the reference momentum in the MV model (left) and after JIMWLK evolution (right). We compare $\ptt=\ptt^{\rm Ref}$ to the case of a wider reference momentum bin $0.5\,\qs < \ptt^{\rm Ref} < 3\, \qs$. Solid lines correspond to results from the lattice simulation without additional approximation, dash-dotted lines (with squares) show results in the non-linear Gaussian approximation, and dotted line (with circles) correspond to the Glasma graph approximation.}
\end{figure*}

In order to establish the quality of the different approximations, we will now compare the results for the Fourier coefficients $v_{n}$ to numerical lattice computations that fully evaluate the correlation functions of Wilson lines. We begin with a comparison in the McLerran-Venugopalan (MV) model where the Wilson lines are generated from a Gaussian ensemble of fluctuating color charges. Following the numerical procedure of Ref.~\cite{Lappi:2007ku}, a set of Wilson line configurations is generated according to \eq\nr{eq:wilsonLine}; these are then employed to extract numerically the expectation value of the dipole operator $D(\rt)$. With the expectation value of the dipole operator we can then compute the double inclusive spectrum in \eq\nr{eq:DoubleInclusive} using the dipole-dipole correlator in the Glasma graph approximation of \eq\nr{eq:SingleGluonExchangev2} and in the nonlinear Gaussian approximation of \eq\nr{eq:NonLinearGaussian}. We also compute the double inclusive spectrum directly from the lattice Wilson lines using the procedure described in Ref.\,\cite{Lappi:2015vha}. For each of these three different double inclusive spectra, we determine the Fourier coefficients $v_n$ using \eqs\nr{eq:FourierDec} and~\nr{eq:v22}. We choose the reference momentum to be $\ptt^{\rm Ref}=\ptt$ such that $v_n(\ptt)=\sqrt{V_{n\Delta}(\ptt)}$.
Our results for the Fourier coefficients $v_{n}(\ptt)$ in the MV model are shown in the left panel of \fig\ref{fig:vNModel}.

We also perform, as discussed in~\cite{Lappi:2015vha},
the JIMWLK rapidity evolution of the Wilson lines for $y=7.6$ units in rapidity 
for SU(3) and for $y=12.4$ units for SU(2) with the same running coupling
formula and  an
initial saturation scale $\qs/\lqcd=3.7$ in both cases. 
We use the running coupling
prescription for the JIMWLK equation proposed in Ref.~\cite{Lappi:2012vw}.
We then compute again the double inclusive spectrum and the Fourier harmonics 
directly from the lattice Wilson lines, as well as 
from the lattice result for the dipole. The  results including JIMWLK evolution are shown in the right panel of \fig\ref{fig:vNModel}.
For the MV model the probe size is  $B\qs^2 = 3.7$ and for the JIMWLK
simulations  $B\qs^2 = 2.5$; for a discussion of the $B$-dependence see~\cite{Lappi:2015vha}. 

 We note that for the MV model case the nonlinear Gaussian approximation agrees perfectly with the direct numerical calculation for all $v_n$ up to the numerically accessible values of  $\ptt$. This is of course a simple  numerical check of the analytical expressions and the agreement should be exact because non-Gaussianities are absent by definition in the MV model. In contrast, the Glasma graph approximation deviates significantly from the numerical result. As discussed previously, the Glasma graph result does not have any odd harmonics. However the even Fourier coefficients $v_{2}$ and $v_4$ too  show significant deviations from the exact numerical result 
especially around $\ptt\sim 2\qs$, demonstrating that nonlinearities are 
significant even at larger $\ptt$. The Glasma graph 
approximation should be exact in the high momentum limit when $|\pt|$,
$|\qt|$, $|\pt-\qt|$  and $|\pt+\qt|$ are all large. However, even for a
large $|\pt| = |\qt|$ the azimuthal harmonics receive contributions
from $|\pt\pm\qt| \lesssim \qs$ where the two-gluon exchange approximation 
is not very accurate.

With JIMWLK rapidity evolution, the distribution of color charges is no longer explicitly Gaussian.
Therefore, although non-Gaussian contributions were not seen in the operators studied previously in~\cite{Dumitru:2011vk}, it is possible
that JIMWLK evolution of the azimuthal anisotropies will introduce non-Gaussian contributions. Indeed this is seen in \fig\ref{fig:vNModel} (right) where we observe a deviation of the nonlinear Gaussian approximation from the numerical result from solving the JIMWLK equations. Furthermore we see that both the numerical and the nonlinear Gaussian results for all $v_n$ are reduced by the JIMWLK evolution. On the contrary, the Glasma graph results, after JIMWLK rapidity evolution of the Wilson lines, are roughly the same as those for the MV model. Thus while better agreement of the Glasma graph approximation with the nonlinear Gaussian and full JIMWLK results is seen after rapidity evolution, this agreement is a fortuitous numerical coincidence.

We have also analyzed the $\nc$ dependence of our results for $v_n$ employing the full numerical calculation of the dipole-dipole correlator. Our results for the SU(2) and SU(3) gauge theory are shown in Fig.\,\ref{fig:vnNc}, where we scale the Fourier coefficients $v_{2}$ and $v_{4}$ by the color factor $\sqrt{\nc^2-1}$. This is because azimuthal correlations in the double inclusive spectrum contain  an overall factor of $1/(\nc^2-1)$ (see e.g. \eq\nr{eq:GGAcoll}) and the $v_n$'s are related via a square root to the Fourier coefficients $V_{n\Delta}$ in the expansion of the double inclusive spectrum. In fact, we find that this scaling works nearly perfectly both in the MV model case (\fig\ref{fig:vnNc} (left)) and the JIMWLK evolved case (\fig\ref{fig:vnNc} (right)); in the former, small differences are seen 
in $v_2$ and $v_4$ for only for large $\ptt$, where lattice cutoff effects
can already have an effect. 

Finally, we demonstrate the dependence of our results on different choices for the reference transverse momentum in Fig.\,\ref{fig:vnpT}. This is an interesting exercise because similar studies can be performed with the experimental data and will help to distinguish between different models. We find a clear suppression of the signal to the previously regarded case $\ptt^{\rm Ref} = \ptt$ when employing a fixed reference bin $0.5\,\qs < \ptt^{\rm Ref} < 3\,\qs$. One observes that for the Gaussian correlations in the MV model, this effect is particularly strong at large $\ptt$ in the Glasma graph approximation. In case of the JIMWLK-evolved results, all approximations show a similarly strong suppression of the signal at large $\ptt$ when using the stated fixed reference momentum bin.

The decorrelation in $\ptt$ observed in \fig\ref{fig:vnpT} is fairly fast and appears incompatible with the experimental observations. Experimentally only a slow decorrelation can be seen in the data when comparing experimental results for $v_n$ in p+Pb collisions using different methods~\cite{Abelev:2012ola,Chatrchyan:2013nka}. However a number of caveats are in order with regard to this comparison. We emphasize that our results are for quarks or more generally on the parton level. While hadronization effects will weaken the strong dependence on the choice of reference momentum observed on the parton level, a quantitative description of the experimental data in initial state frameworks will also be quite sensitive to the choice of the fragmentation scheme. The role of fragmentation in such correlations deserves a more detailed study in the future (see 
also~\cite{Zhou:2015iba,Esposito:2015yva,Ortiz:2015cma}).

\section{The CGC and the color field domain model}

We will now discuss the relation of the azimuthal correlations derived in the CGC framework to those computed recently in the color field domain model introduced in~\cite{Dumitru:2014dra,Dumitru:2014yza,Dumitru:2014vka,Skokov:2014tka}. Since the latter qualitatively describes some key features of the ridge data in proton-lead collisions at the LHC, it is interesting to compare and contrast this model with the CGC based calculations we discussed thus far. 

\subsection{Electric fields in the Glasma graph approximation}

The color field domain model is usually formulated in terms of transverse color electric fields and their correlators. 
To achieve an ``apples-to-apples'' comparison,  we will first show how our previous discussion in terms of dipole-dipole correlators can be formulated in terms of color electric fields. 

The classical color electric fields in the target, as shown in Fig.~\ref{fig:ColorFields}, can be expressed in terms of the Wilson line correlators as 
 \begin{equation}
 \label{eq:diluteE}
E_i(\xt) = i V(\xt)\partial_{i}V^{\dagger}(\xt)\;.
\end{equation}
Performing a short distance expansion of the dipole operator, one obtains 
\begin{align}
\label{eq:E-dipole}
\mathcal{D}(\xt,\yt) \simeq 1 -  \frac{r^i r^j}{4\nc} E_i^a(\bt) E_j^a(\bt).
\end{align}
where we denote $\rt=\xt-\yt$ and $\bt=(\xt+\yt)/2$. Within the weak field limit of the Glasma graph approximation outlined previously, one can evaluate the correlator of color electric fields as
\begin{equation}
\label{eq:E2Standard}
\left< E^i_a(\xt)E^j_b(\yt) \right>= -\delta^{ab} \partial^{i}\partial^{j} \gamma(\xt-\yt) \, , 
\end{equation}
where we have used Eq.~(\ref{eq:defgamma}). We emphasize however that this equivalence is valid only in the combined limit of weak fields and short distances. 
Using the electric field correlator then yields the following expression for the dipole operator
\begin{eqnarray}
\langle \mathcal{D}(\xt,\yt)\rangle \simeq 1+ \cf  \frac{r^{i}r^{j}}{2} \partial_{i}\partial_{j} \left. \gamma(\rt) \right|_{\rt=\ot} \; .
\label{eq:dipoleQs}
\end{eqnarray}

One can similarly use Eq.~(\ref{eq:E-dipole}) and express the dipole-dipole correlator in the short distance limit as 
\begin{multline}
\label{eq:DDfromEE}
\langle \mathcal{D}(\xt,\yt) \mathcal{D}(\ut,\vt) \rangle 
\simeq D(\xt-\yt) D(\ut-\vt) 
 \\
+ \frac{\rt^i_{1} \rt^j_{1} \rt^k_{2} \rt^l_{2} }{16 \nc^2} \Bigg[ \Big<E_i^a(\bt_1) E_j^a(\bt_1) E_k^b(\bt_2) E_l^b(\bt_2)\Big> 
\\
 -\Big<E_i^a(\bt_1) E_j^a(\bt_1)\Big> \Big<E_k^b(\bt_2) E_l^b(\bt_2)\Big> \Bigg]\;.
\end{multline}
where $\xt=\bt_1+\rt_1/2,\yt=\bt_1-\rt_1/2,\ut=\bt_2+\rt_2/2,\vt=\bt_2-\rt_2/2$. This expression shows that the expectation value of the dipole-dipole correlator is sensitive to fluctuations of the color electric fields as characterized by the four point correlator.

In the Glasma graph approximation, the electric field is linearly proportional
to the charge density and thus has Gaussian correlations
\begin{eqnarray}
\label{eq:E4Gauss}
&&\left< E^i_a(\xt) E^j_b(\yt)E^k_c(\ut)E^l_d(\vt) \right> = \\
&&\qquad \qquad \qquad  ~\Big< E^i_a(\xt) E^j_b(\yt) \Big> 
\Big< E^k_c(\ut)E^l_d(\vt) \Big> \nonumber \\
&&\qquad \qquad \qquad +\Big< E^i_a(\xt) E^k_c(\ut) \Big> 
\Big< E^j_b(\yt)E^l_d(\vt) \Big> \nonumber \\
&&\qquad \qquad \qquad  +\Big< E^i_a(\xt) E^l_d(\vt) \Big> 
\Big< E^k_c(\ut)E^j_b(\yt) \Big>\,. \nonumber 
\end{eqnarray}
The four point correlation function can be expressed in terms of the two point function as
\begin{eqnarray}
\label{eq:mv4ptfinal}
&&\left< E^i_a(\xt) E^j_b(\yt)E^k_c(\ut)E^l_d(\vt) \right> =\\
&&~\delta^{ab} \delta^{cd}\partial_i\partial_j \gamma(\xt-\yt)\partial_k\partial_l \gamma(\ut-\vt) \nonumber \\
&&+\delta^{ac} \delta^{bd}
\partial_i\partial_k \gamma(\xt-\ut)\partial_j\partial_l \gamma(\yt-\vt) \nonumber \\
&&+\delta^{ad} \delta^{bc}
\partial_i\partial_l \gamma(\xt-\vt)\partial_j\partial_k \gamma(\yt-\ut) \nonumber.
\end{eqnarray}
Combining the above expressions, the dipole-dipole correlator takes the form
\begin{multline}
\langle \mathcal{D}(\xt,\yt) \mathcal{D}(\ut,\vt) \rangle  \simeq D(\xt-\yt) D(\ut-\vt) 
\\
 +\frac{\cf^2}{2(\nc^2-1)} \Big( \rt_{1}^{i} \rt_{2}^{j} \partial_{i} \partial_{j} \gamma(\bt_{1}-\bt_{2})\Big)^2,
\label{eq:shortdistgaus}
\end{multline}
which agrees precisely with the expansion of the result in the double gluon exchange approximation in 
\eq\nr{eq:SingleGluonExchange} in the short distance limit $|\rt_1|\sim |\rt_2| \ll |\bt_1| \sim |\bt_2|$.  \\

This simple calculation shows that the physics of fluctuating color electric field domains is implicitly contained in the conventional Glasma graph picture. While in the short distance (large momentum) limit  the dipole-dipole correlator is expressed in terms of two and four point correlators of electric fields, there is a one-to-one mapping
between the statistical properties of these electric fields and those of the $\Lambda$'s in the Glasma graph calculation. 

\subsection{The color field domain model and non-Gaussian correlations}

We focused thus far on conventional models based on Gaussian correlations of color fields 
inside a large nucleus. It is now interesting to understand how these relate to the color field domain model~\cite{Dumitru:2014dra,Dumitru:2014yza,Dumitru:2014vka,Skokov:2014tka}. In our language, the color field domain model is obtained by replacing the electric field correlator in Eq.~\nr{eq:E2Standard} by
\begin{eqnarray}
\label{eq:ModifiedE}
\left< E^i_a(\xt)E^j_b(\yt) \right>&=& -\frac{\delta^{ab}}{2} \left[\delta^{ij}(1-\A)+ 2\A \ahat_i\ahat_j \right] \nonumber \\
&& \times \nabt^2 \gamma(\xt-\yt) \, .
\end{eqnarray}
This correlator in the color field domain model depends explicitly on the effective degree of polarization $\A$ and 
the unit vector $\ahat$ characterizing the direction of the color electric field. 

Expectation values of operators within the color field domain model are computed by a two step averaging procedure.  In the first step, one performs a Gaussian average with the modified two point correlation function in Eq.~(\ref{eq:ModifiedE}). The second step consists of an average over all possible directions of the chromoelectric fields, $\ahat$, such that
\begin{eqnarray} 
\label{eq:ahat1}
\left< \ahat^i \ahat^j \right>_\ahat &=& \frac{1}{2}\delta^{ij}\;,
\end{eqnarray}
and 
\begin{eqnarray}
\label{eq:ahat2}
&&\left<\ahat^i \ahat^j\ahat^k \ahat^l \right>_\ahat = \frac{\delta^{ij}\delta^{kl}+\delta^{ik}\delta^{jl} +\delta^{il}\delta^{jk}}{8}\;. \nonumber
\end{eqnarray}

Implicit in this two step procedure is a physical assumption about time scales. One assumes that partons within a color domain of size $\sim1/\qs$ generate a color electric field oriented in a particular direction $\ahat$ with a likelihood  $\A$ ranging from $0-100\%$ , which is long lived on the time scale of the interaction such that partons in the projectile are collimated relative to the direction $\ahat$ of this color electric field. A similar picture is implicit in the work of \cite{Gyulassy:2014cfa} where the net momentum transfer from the projectile partons to the target takes the place of the color electric field. 

When the life time of the degrees of freedom responsible for the breaking of rotational symmetry within each color domain is much larger than the time scale over which one performs the average over the color electric fields inside the target in \eq\nr{eq:ModifiedE}, the orientation of the domain appears frozen on that time scale and a separate averaging is justifiable.  However it is not \emph{a priori} evident that such a separation of time scales exists. In particular, it is not clear what would be the intrinsic or dynamical scale that separates the time scale over which the color electric fields align themselves from the time scale over which one performs the average over the different orientations of the field and why such an average should be a Gaussian average.

Let us now discuss explicitly the calculation of the four point 
correlator of color electric fields in the color field domain 
model. The only correlators that are related to physical 
observables are the ones 
averaged over all unobservable degrees of freedom, including the 
direction of $\ahat$. Therefore, to understand the correlation 
structure of the model we 
must compare the two- and four-point functions of the electric 
field after the full two step average.

Carrying out the two step averaging procedure for two point correlators by averaging \eq\nr{eq:ModifiedE} over $\ahat$ using \eq\nr{eq:ahat1} one obtains
\begin{eqnarray}
 \label{eq:aterm2pt}
\left<\left< E^i_a(\xt)E^j_b(\yt) \right> \right>_\ahat=\frac{\delta^{ab}}{2}\delta^{ij} \nabt^2 \gamma(\xt-\yt).
\end{eqnarray}
The result is independent of the effective polarization $\A$ and the normalization has been chosen in such a way that the two-point function \eq\nr{eq:aterm2pt} agrees with the previous result in \eq\nr{eq:E2Standard} for a rotationally invariant system in the short distance limit\footnote{Note that for a rotationally invariant correlation function $\gamma(\rt)$ in the short distance limit $r \to 0$ we can replace
\begin{equation}
\partial_i\partial_j \gamma(\rt)|_{\rt=\ot}
= \frac{\delta^{ij}}{2}\nabt^2 \gamma(\rt)|_{\rt=\ot}.
\end{equation}
}.

The short distance expansion of the dipole-dipole correlator in \eq\nr{eq:DDfromEE} involves the (double) average of the four point correlation function which, after the first step, can be expressed as 
\begin{multline}
\label{eq:aterm4pt}
\left<\left< E^i_a(\xt) E^j_b(\yt)E^k_c(\ut)E^l_d(\vt) \right> \right>_\ahat=
\\
\Bigg< \Big< E^i_a(\xt) E^j_b(\yt) \Big> \Big< E^k_c(\ut)E^l_d(\vt) \Big> 
\\
+\Big< E^i_a(\xt) E^k_c(\ut) \Big> \Big< E^j_b(\yt)E^l_d(\vt) \Big> 
\\
+\Big< E^i_a(\xt) E^l_d(\vt) \Big> \Big< E^k_c(\ut)E^j_b(\yt) \Big> \Bigg>_\ahat.
\end{multline}
Performing the second average with Eq.~(\ref{eq:ahat2}), one obtains 
\begin{align}
\label{eq:aterm4ptfinal}
&
\left<\left< E^i_a(\xt) E^j_b(\yt)E^k_c(\ut)E^l_d(\vt) \right> \right>_\ahat=
\\  \nonumber 
& \qquad
\frac{\delta^{ab}\delta^{cd}}{4}\left[
\left(1-\frac{1}{2}\A^2\right) 
\delta^{ij}\delta^{kl}
+ \frac{\A^2}{2}\left(\delta^{ik}\delta^{jl}+\delta^{il}\delta^{jk}\right)
\right]
\\ \nonumber
& \qquad \qquad \qquad \times
 \nabt^2\gamma(\xt-\yt)  \nabt^2\gamma(\ut-\vt)
\\ \nonumber
&\quad +
\frac{\delta^{ac}\delta^{bd}}{4}\left[
\left(1-\frac{1}{2}\A^2\right) 
\delta^{ik}\delta^{jl}
+ \frac{\A^2}{2}\left(\delta^{ij}\delta^{kl}+\delta^{il}\delta^{jk}\right)
\right]
\\ \nonumber
& \qquad \qquad \qquad \times
 \nabt^2\gamma(\xt-\ut) 
 \nabt^2\gamma(\yt-\vt) 
\\ \nonumber
&\quad +
\frac{\delta^{ad}\delta^{bc}}{4}\left[
\left(1-\frac{1}{2}\A^2\right) 
\delta^{il}\delta^{jk}
+ \frac{\A^2}{2}\left(\delta^{ij}\delta^{kl} + \delta^{ik}\delta^{jl} \right)
\right]
\\ \nonumber
& \qquad \qquad \qquad \times
 \nabt^2\gamma(\xt-\vt) 
 \nabt^2\gamma(\yt-\ut) .
\end{align}
Comparing the two point and four point correlation functions of the color electric fields in \eqs\nr{eq:aterm2pt} and~\nr{eq:aterm4ptfinal}, one sees explicitly that the four point correlation function can not be expressed in terms of the two point function, i.e. the
color field domain model is non-Gaussian. 
Specifically one finds that the dipole-dipole correlator
\begin{align} \label{eq:nonGaussDD}
&\!\!\!\!\!\!\!\!\! \langle \mathcal{D}(\xt_{1},\yt_{1}) \mathcal{D}(\xt_{2},\yt_{2}) \rangle \simeq  D(\rt_1)D(\rt_2)  \nonumber  \\
&+~\frac{ \cf^2}{8(\nc^2-1)} \Bigg( (\rt_{1} \cdot \rt_{2}) \nabt^2 \gamma(\bt_1-\bt_2)\Bigg)^2 \nonumber \\
& +~\frac{\cf^2 \A^2}{32}  \Big( 2(\rt_{1}\cdot \rt_{2})^2-\rt_{1}^2\rt_{2}^2 \Big)   \Bigg(\nabt^2 \gamma(\ot) \Bigg)^2 \nonumber \\
&  +~\frac{\cf^2  \A^2}{16(\nc^2-1)} \rt_{1}^2\rt_{2}^2 \Bigg(\nabt^2 \gamma(\bt_1-\bt_2) \Bigg)^2
\end{align}
is a sum of a Gaussian piece (present already in \eq\nr{eq:shortdistgaus})
and non-Gaussian terms proportional to $\A^2$ induced by the two step averaging 
procedure. The non-Gaussian terms are referred to as ``disconnected 
contributions'' in Ref.~\cite{Dumitru:2014yza}.
In addition to the small dipole limit  $\rt_1^2,\rt_2^2 \ll 1/\qs^2$
that is assumed in this discussion, one can additionally take the 
limit $\bt_1^2,\bt_2^2 \ll 1/\qs^2$. 
 In this case the two-gluon exchange  approximation becomes exact and 
thus the Glasma graphs and the nonlinear
Gaussian are equivalent to each other. 
Even in this limit the $\A$-terms are not suppressed in any way 
and the color field domain model remains different from the other
approaches considered in this paper.

It is obvious that the $(\rt_1\cdot \rt_2)^2$ terms introduce an additional $\A$-dependent $\cos 2 \phi$ correlation between the two dipoles. Upon Fourier transformation, 
this modifies the angular structure of the correlation between the two produced particles. What remains unclear at this stage is the physical origin of this particular
form of non-Gaussian correlators as well as the magnitude of the non Gaussianity characterized by the additional $\A$ parameter in this model.

\subsection{Interpretations of the color field domain model}

We noted previously but wish to emphasize again that fluctuating domains of color electric field are present also in the Glasma graph or non-linear Gaussian approximation and they are not a new physical feature added by the color field domain model. What is different in the color field domain model is that the direction of the chromoelectric field is treated explicitly as a long lived degree of freedom. By modifying the correlation function of electric fields according to Eq.~\nr{eq:ModifiedE} and performing a separate average over the orientation of the electric fields the statistics of these domains is altered significantly. Most importantly, this can lead to sizable non-Gaussian correlations depending on the magnitude of the parameter $\A$ in this model. Since the single inclusive distribution is not sensitive to $\A$, its value can only be determined from correlation measurements. 
An observable that would be particularly sensitive to the presence of
intrinsic non-Gaussianities would be the four-particle cumulant flow 
coefficient, as discussed in~\cite{Dumitru:2014yza}.
We will now discuss three possible interpretations of the non-Gaussian correlations represented by the $\A$-term in the color domain model.

\begin{enumerate}
\item 
The color electric field is a nonlinear function of the color charge density.
Thus even if the color charges (or, equivalently the $\Lambda$'s) have a
Gaussian distribution, the color electric fields
and thus the dipole operators can have non-Gaussian correlations. 
One possible interpretation of
the color field domain model is as an effective way to account  for the
non-linear relation between electric fields and color charge densities in 
the target in an otherwise linearized calculation. 
The Glasma graph approximation of Eq.~\nr{eq:diluteE} assumes that the 
electric field is linearly proportional to the color charge and thus  
the electric fields have Gaussian correlations.
The nonlinear Gaussian approximation, on the
other hand, has Gaussian correlations for color charges but not for the electric fields. In this interpretation of the color field domain 
model, the corrections encoded in $\A$ should be proportional to the 
difference between the Glasma graph and the nonlinear Gaussian computations in 
this paper.
The total anisotropy of the azimuthal two-particle distribution 
calculated from the MV model (see e.g.~\cite{Dumitru:2014vka}),
is then the sum of the Glasma graphs and the $\A$-term.
In this interpretation one has parametrically $\A^2\sim 1/(\nc^2-1)$, since correlations are $\nc$-suppressed (relative to the uncorrelated term) in both the nonlinear Gaussian and the Glasma graph approximations. Our numerical results in Fig.~\ref{fig:vNModel} show that such non-linear corrections can indeed be sizeable and should be taken into account as a correction to the Glasma graph result. However, if this is the interpretation it would seem more natural to directly use the non-linear Gaussian approximation rather than introducing an additional parameter. In particular, it is not obvious whether a constant
$\A$ could have a similar momentum dependence as the nonlinear Gaussian 
approximation, given that the color field domain model differes from the 
nonlinear Gaussian even in the small distance limit.
\item A second possible interpretation of the $\A$-term is that it represents non-Gaussian
correlations that can emerge from JIMWLK rapidity evolution even when starting
from a Gaussian initial condition. This contribution is, in our 
present calculation, represented by the difference between the
full JIMWLK result and the nonlinear Gaussian. Indeed we see signs
of a $\sim 10\%$ deviation between the two for $\ptt \gtrsim \qs$. 
However, this difference is relatively small in practical terms and might not have a significant influence on phenomenology. 
From a theory perspective, it is to our knowledge the first instance 
observed in the literature of a
meaningful breaking of the Gaussian approximation to JIMWLK.
We see no obvious reason why the deviation from Gaussianity seems larger
here than in the observables studied in Ref.~\cite{Dumitru:2011vk}. 
This issue might call for additional studies in the future including a more systematical check of discretization effects in the lattice calculations.
\item Finally the most intriguing possibility is that the $\A$-term represents an intrinsic non-Gaussian
correlation that is present in the initial condition for JIMWLK evolution and survives substantially after evolution.
This is the interpretation suggested in~\cite{Dumitru:2014dra,Dumitru:2014yza}.
The possible existence of such non-Gaussian correlators was previously suggested in \cite{Kovner:2010xk,Kovner:2011pe}
and later studied in \cite{Dumitru:2011ax,Dumitru:2011zz}. While the Gaussian MV model can be justified on quite general  grounds~\cite{Jeon:2004rk} as arising, due to the central limit theorem, from a superposition of a large amount of uncorrelated color charges in a heavy nucleus, deviations from Gaussian statistics are naturally expected for a small number of large $x$ degrees of freedom. The existence and persistence of 
such a non-Gaussianity at small $x$ would thus be a signal of remarkably strong long range rapidity correlations inside the gluon cascade building up the strong color fields at small $x$. 
The computations in this paper do not address this possibility of 
an intrinsic non-Gaussian four-particle correlation because we 
have been  working in the MV model+JIMWLK evolution setup where 
such correlations are absent in the initial conditions.
\end{enumerate}

\section{Summary and conclusions}
We explored a number of  calculational schemes to compute two particle correlations of quarks scattering off a highly energetic nucleus.
All cases correspond to different approximations within the dilute-dense limit of the color glass condensate framework.
The two-particle correlations are quantified in terms of Fourier coefficients in an expansion in relative azimuthal angle of the double inclusive distribution of scattered quarks. This distribution is proportional to the dipole-dipole correlator. The study of the properties of this correlator in the various approximation schemes was the primary objective of this work. 

The simplest approximation scheme considered was the glasma graph approximation. In this case, the lightlike Wilson lines in the dipole-dipole correlator are expanded to lowest order, restricting the interaction with the target to two gluon exchange. One further assumes that the gluon correlations in the target are Gaussian correlations. This approximation scheme has been used previously in the literature to study  azimuthally collimated double inclusive gluon production in p+p and p+Pb collisions.

Another approximation scheme, of greater complexity, is the nonlinear Gaussian approximation. In this case, all multigluon exchanges are resummed to all orders to obtain a complicated analytical expression for the dipole-dipole correlator.  This expression is exact as long as there are only Gaussian correlations in the target. This is for instance the case for the MV model.
 
These analytical results are a good benchmark for numerical studies wherein the Wilson lines are computed on 2+1-dimensional lattices for Gaussian distributed sources - good agreement is expected and achieved. With the MV initial conditions for the Wilson lines at a given rapidity, the JIMWLK equations are solved on the lattice to determine the Wilson lines at larger rapidities. These then allow one to determine in principle expectation values of n-dipole correlators as a function of rapidity.

We studied the Fourier harmonics $v_2$, $v_3$, and $v_4$  that are extracted from azimuthal two particle correlations in the various approximation schemes. The Glasma graph approximation 
and the nonlinear Gaussian approximations differ appreciably for $\ptt\sim 2 \qs$ in the MV model, indicating the importance of
coherent multiple scattering effects. 
Since the Glasma graph approximation is at the heart of most comparisons to experimental data, this calls for a more detailed study to further quantify its theoretical uncertainties.
However, we believe that in terms of the phenomenological consequences most of this difference can be accomodated within the uncertainties in the overall normalization of the Glasma graph calculations~\cite{Dumitru:2010iy,Dusling:2012iga,Dusling:2012cg,Dusling:2012wy,Dusling:2013oia}. 
A significant difference between the two approximation schemes is that the symmetry constraints inherent in the Glasma graph approximation do not produce any odd harmonics in contrast to the nonlinear Gaussian approximation which generates all odd harmonics of the azimuthal double inclusive distribution.  With JIMWLK rapidity evolution, the differences in the $v_{2,4}$ coefficients computed in the Glasma graph and nonlinear Gaussian schemes decrease significantly. 
Furthermore, we find that the coefficients in the two schemes are quite close to those computed by solving JIMWLK numerically without any approximation to the dipole-dipole correlator. Since we do not currently have a good interpretation for this better agreement, we believe that it may to some extent be accidental.

We analyzed the dependence of our results on the number of colors by comparing computations for SU(3) and SU(2) gauge fields and found precisely the expected scaling of $V_{n\Delta}(\ptt)$ with $1/(\nc^2-1)$. This result confirmed that the azimuthal angle dependent correlations are suppressed parametrically by $1/\nc^2$.

We studied the dependence of the Fourier coefficients on the reference transverse momentum $\ptt^{\rm Ref}$ (the momentum of the second scattered quark) in the different approximation schemes. These showed clear differences for the two reference momenta considered: $\ptt^{\rm Ref}=\ptt$, and $0.5\,Q_s<\ptt^{\rm Ref} <3\,Q_s$. For all the approximation schemes, the choice of  equal $\ptt$ led to a larger signal for $\ptt\gtrsim Q_s$, with the Glasma graph approximation showing the largest differences. JIMWLK rapidity evolution seems to increase the difference between the different $\ptt^{\rm Ref}$ choices. We note however, that the choice of the fragmentation scheme can qualitatively influence the comparison of model computations of gluon correlations to the hadron correlation data.  This topic deserves a more
detailed study in the future.

Finally we analyzed in detail the relation of the color glass condensate based computations to a color domain model which captures qualitative features of the multiparticle azimuthal correlations observed in proton-nucleus collisions. In this framework, the dipole-dipole correlator is modified to include an additional term that models the polarization of gluon fields in individual domains of color charge within the target. We conclude that this term, proportional to the polarization parameter $\A$, introduces non-Gaussian correlations amongst the color electric fields inside the target nucleus. On 
the other hand, if such non-Gaussianities are not explicitly introduced, 
the color domain model reduces to the MV model in the Glasma graph approximation.

We also discussed possible origins of non-Gaussian correlations of the color
fields of a large nucleus.
One possibility is that these correspond to non-Gaussian correlations induced by JIMWLK rapidity evolution of Gaussian correlations at the initial rapidity. However our estimates of this effect suggest that such non-Gaussian correlations are too small to be relevant phenomenologically. A more interesting possibility is that the $\A$ polarization term introduced in this model arises from intrinsic four point correlations that are significant in the initial condition and whose magnitude is preserved with rapidity evolution. 
Such correlations would be interesting to study in the future.

\section*{Acknowledgments}
We would like to thank A. Dumitru, A. Kovner and V. Skokov for useful 
discussions.
T.~L.\ is supported by the Academy of Finland, projects
267321 and 273464.
BPS, SS, and RV are supported under DOE Contract No. DE-SC0012704. This research used computing resources of 
CSC -- IT Center for Science in Espoo, Finland and of
the National Energy Research
Scientific Computing Center, which is supported by the Office of Science of the U.S. Department of Energy under Contract No. DE-AC02-05CH11231. RV would like to thank the Institut f\"{u}r Theoretische Physik, Heidelberg, for kind hospitality and the Excellence Initiative of Heidelberg University for their support. SS gratefully acknowledges a Goldhaber Distinguished Fellowship from Brookhaven Science Associates. BPS is supported by a DOE Office of Science Early Career Award.
TL thanks the BNL for hospitality during this work. 

\bibliography{spires}
\bibliographystyle{JHEP-2modlong}

\end{document}